\documentclass{siamart1116}
\usepackage{bm}
\usepackage{bbm}
\usepackage{graphicx}
\usepackage{url}
\usepackage{color}
\usepackage{mathtools}
\usepackage{amsmath}	
\usepackage{amssymb}	
\usepackage{graphicx,graphics,color,subcaption}	
\usepackage{enumerate}
\usepackage{cancel}
\usepackage{color}
\usepackage{cite}

\newtheorem{thm}{Theorem}[section] 
\newtheorem{remark}[thm]{Remark}

\newcommand{\drt}[1]{{\color{black}#1}}

\DeclareMathOperator{\Tr}{Tr}

\title{Network-Ensemble Comparisons with Stochastic Rewiring and Von Neumann Entropy
\thanks{This work was funded in part by NIH Grant No.~R01HD075712 and James S. McDonnell Grant  \#220020315.
Any opinions, findings, and conclusions or recommendations expressed in this material are those of the authors and do not necessarily reflect the views of the funding agencies. }}
\author{
Zichao Li
\footnotemark[3]
\thanks{Department of Statistics and Operations Research, University of North Carolina, Chapel Hill, NC 27599, USA (lizichao@live.unc.edu)}
\and
Peter J. Mucha
\thanks{Carolina Center for Interdisciplinary Applied Mathematics, Department of Mathematics, University of North Carolina, Chapel Hill, NC 27599, USA (mucha@unc.edu)}
\and
Dane Taylor\footnotemark[3]
\thanks{Department of Mathematics, University at Buffalo, State University of New York (SUNY), Buffalo, NY 14260, USA (danet@buffalo.edu})
}

\begin{document}
\maketitle

\begin{abstract}
Assessing whether a given network is typical or atypical for a random-network ensemble (i.e., \emph{network-ensemble comparison}) has widespread applications ranging from null-model selection and hypothesis testing to clustering and classifying networks. We develop a framework for network-ensemble comparison by subjecting the network to stochastic rewiring. We study two rewiring processes---uniform and degree-preserved rewiring---which yield random-network ensembles that converge to the Erd\H{o}s-R\'enyi and configuration-model ensembles, respectively. We study convergence through von Neumann entropy (VNE)---a network summary statistic measuring information content based on the spectra of a Laplacian matrix---and develop a perturbation analysis for the expected effect of rewiring on VNE. Our analysis yields an estimate for how many rewires are required for a given network to resemble a typical network from an ensemble, offering a computationally efficient quantity for network-ensemble comparison that does not require simulation of the corresponding rewiring process.
\end{abstract}



\pagestyle{myheadings}
\thispagestyle{plain}
\markboth{Z. Li, P. J. Mucha, and D. Taylor}{Network-Ensemble Comparisons with Rewiring}

\section{Introduction}

Numerous social, biological, technological and information systems are naturally manifest as networks \cite{newman2010networks}, and common questions about networks are cast explicitly or relate implicitly to network models and their corresponding network ensembles---a set of networks combined with a sampling probability distribution.  The Erd\H{o}s-R\'enyi (ER) and configuration model random-network ensembles, for example, have provided cornerstones for the development of graph theory \cite{erdos1959random,bollobas2013modern,chung1997spectral} and are widely used as null models for network-data analytics including community detection
 \cite{mucha2010community,newman2004finding} and significance testing of subgraph motifs \cite{milo2002network}.
Moreover, many applications call for the simultaneous study of a set of empirical networks, encoded as layers in a multilayer (e.g., multiplex) network \cite{boccaletti2014structure,kivela2014multilayer}, where it can be beneficial to study them as independent samples from an ensemble \cite{stanley2015clustering,taylor2016enhanced,taylor2016detectability}.

We pursue here two classes of questions related to network ensembles: \emph{network-network comparison} and \emph{network-ensemble comparison}. Network-network comparison aims to identify a similarity measure between networks, for instance as a means for clustering and classifying networks \cite{sizemore2015classification,aliakbary2015distance,gallos2014revealing,monnig2016resistance,soundarajan2014guide,berlingerio2013network,papadimitriou2010web,koutra2013d,macindoe2010graph,jurman2015him,de2015structural,de2016spectral,onnela2012taxonomies,caceres2016model}.
Closely-related questions of network-ensemble comparison aim to assess whether a given network is typical or atypical for an ensemble (or to quantify how typical).  Such comparison is useful for null-model selection and hypothesis testing \cite{bassett2013robust,caceres2016model,kolaczyk2014statistical,milo2002network}. Understanding if a given network is typical for an  ensemble is particularly important for modeling dynamics on networks (e.g., epidemic spreading \cite{pastor2001epidemic}, social contagions \cite{gleeson2013binary}, synchronization \cite{restrepo2005onset,skardal2013effects}, neuronal excitation \cite{larremore2011predicting}, percolation theory \cite{cohen2000resilience,taylor2012network}, and so on). Specifically, the accuracies of mean field theories or other model system reductions to describe the expected dynamics for random-network ensembles are related to whether a network is typical or atypical for an ensemble\cite{melnik2011unreasonable,gleeson2012accuracy,taylor2012network,radicchi2015predicting}.

We study network-network and network-ensemble comparisons through von Neumann entropy (VNE), a summary statistic that measures a network's information content based on the spectra of its associated Laplacian matrix \cite{braunstein2006laplacian,passerini2008neumann}. VNE-based comparison is closely related to the family of network-network comparisons known as spectral comparisons \cite{jurman2011introduction}, which relate networks by some function of the eigenvalues of matrices associated with the networks (e.g., adjacency, normalized Laplacian, and unnormalized Laplacian).
Our focus on VNE is motivated by recent research \cite{de2016spectral,de2015structural} that used VNE to hierarchically cluster layers in multilayer networks. We stress, however, that the mathematical techniques that we develop here can be generalized to other summary statistics of networks.

Our main goal is to study VNE for networks undergoing stochastic rewiring. We study two rewiring processes---uniform and degree-preserved rewiring---that yield random-network ensembles that converge in distribution to the ER and configuration-model ensembles, respectively.  This convergence follows from studying  rewiring as a degree-regular Markov chain in which states represent networks and transitions represent stochastic rewiring. Indeed, stochastic rewiring is an established approach for Markov chain Monte Carlo (MCMC)  algorithms for sampling configuration ensembles \cite{fosdick2016configuring,blitzstein2011sequential,milo2003uniform}. Because stochastic rewiring is an important generative model for time-varying networks \cite{holme2012temporal}, our theory also provides insight about the VNE of time-varying networks. We conduct a perturbation analysis for the change in VNE incurred by rewiring a small number of edges. We prove that the distribution of network summary statistics (e.g., VNE) for an ensemble of networks obtained by rewiring $t$ edges converges as $t \to \infty$ to the appropriate distribution for the associated random-network ensemble. Combining these two results, we obtain \drt{an exponential} extrapolation $B_\alpha$ that predicts how many rewires are necessary to modify an empirical network so that it resembles a typical network from an ensemble. Importantly,  the calculation of $B_\alpha$ does not require the \drt{full} simulation of the stochastic rewiring process, nor the calculation of VNE for rewired networks, and is therefore a computationally efficient quantity for evaluating network-ensemble comparisons.

The remainder of this paper is organized as follows. 
In Sec.~\ref{sec:Back}, we provide background information.
In Sec~\ref{sec:Main}, we present our main mathematical results regarding the VNE of networks undergoing stochastic rewiring.
In Sec.~\ref{sec:Num}, we present numerical experiments and introduce the quantity $B_\alpha$ for network-ensemble comparisons.
We provide a discussion in Sec.~\ref{sec:Discuss}.

\section{Background Information}\label{sec:Back}
We now introduce our mathematical notation and provide background information about the Laplacian matrix (Sec.~\ref{sec:Laplacian}), VNE (Sec.~\ref{sec:VNE}), random-network ensembles (Sec.~\ref{sec:Ensemble}), and Markov-chain theory for stochastic rewiring (Sec.~\ref{sec:Markov}).

\subsection{Laplacian Matrix of a Network}\label{sec:Laplacian}

Let ${G}(\mathcal{E},\mathcal{V})$ denote a network with set $\mathcal{V}=\{1,\dots,N\}$ containing $N=|\mathcal{V}|$ nodes and set $\mathcal{E}\in\mathcal{V}\times\mathcal{V}$ containing $M=|\mathcal{E}|$  edges. We assume the network is simple, unweighted, undirected, and that there are no self-edges. 
The network can be equivalently defined by a symmetric adjacency matrix
\begin{equation} \label{eq:Adj}
A_{ij} = \begin{cases}  
1, & (i,j)\in\mathcal{E} \\
0, &\mbox{otherwise}.
\end{cases}
\end{equation}
Note that $\sum_{ij} A_{ij} = 2M$ since each of the $M$ edges gives rise to two nonzero entries in $A$. We define the degree matrix to be $D=\mbox{diag}[d_1,\dots,d_N]$, where $d_i = \sum_j A_{ij}$ is the degree for each node $i$. The unnormalized Laplacian matrix is given by
\begin{equation} \label{eq:Laplacian}
L = D - A.
\end{equation}
The matrix $L$, also known as the combinatorial Laplacian, is important in numerous applications including graph partitioning \cite{fiedler1989laplacian}, spanning tree analysis \cite{maurer1976matrix}, synchronization of nonlinear dynamical systems \cite{pecora1998master,skardal2014optimal,taylor2016synchronization}, diffusion of random walks \cite{bollobas2013modern,lovasz1993random},  manifold learning \cite{belkin2001laplacian,coifman2006diffusion}, and harmonic analysis \cite{coifman2006harmonic}. In Sec.~\ref{sec:Expected}, we analyze  the expected effect of rewiring on VNE, which requires us to first study the expected effect on $L$. We expect our mathematical results to also potentially benefit these other diverse applications that rely on $L$.

\subsection{von Neumann Entropy (VNE)}\label{sec:VNE}

VNE was  introduced by John von Neumann as a measure for quantum information \cite{neumann2013mathematische} and can quantify, for example, the departure of a quantum-mechanical system from its pure state. Recently, this formalism has been generalized to study information content in networks.

\begin{definition}[von Neumann Entropy of a Network \cite{braunstein2006laplacian}]\label{def:VNE}
Let $ {G}(\mathcal{E},\mathcal{V})$ denote a network, $L$ denote its associated unnormalized Laplacian defined by Eq.~\eqref{eq:Laplacian}, and $\mathcal{L} =  L  / 2M$,
 where $M$ is the number of undirected edges. The VNE of $\drt{G}(\mathcal{E},\mathcal{V})$ is given by
\begin{equation} \label{eq:vne1}
h(G) = -\Tr(\mathcal{L} \log_2 \mathcal{L}) .
\end{equation}
\end{definition}

\begin{remark}
Since $\mathcal{L}$ is positive semi-definite and $\Tr(\mathcal{L}) = \sum_i d_i/2M = 1$, $h$ can be written in terms of the set $\{\lambda_1, \lambda_2, \dots, \lambda_N \}$ of eigenvalues of $L$ as 
\begin{equation} \label{eq:vne2}
h(G) = - \sum_{i = 1}^{N} \frac{\lambda_i}{2M} \log_2\left(\frac{\lambda_i}{2M}\right)
\end{equation}
\drt{[By} convention we define $0 \log_2(0) = 0$]. Note also that because $\Tr(\mathcal{L}) = 1$,  the variables $\lambda_i/2M$ may be interpreted as probabilities. 
\drt{It's worth noting, however, that although $\mathcal{L}$ shares the mathematical properties of a density matrix (i.e., it's positive semidefinite and $\Tr(\mathcal{L})=1$), it does not have the same physical meaning as a density matrix in quantum mechanics.}
\end{remark}
\begin{remark}
Recently, Ref.~\cite{de2016spectral} introduced an alternative notion of VNE for networks using density matrix $\mathcal{L} \varpropto e^{-\beta L}$ for $\beta >0$. They showed this version satisfies the subadditivity property, which can be preferable for some applications, but we will not consider this version further in the present work.
\end{remark}

VNE quantifies the information content of a network through the eigenvalues of its associated Laplacian matrix, which are well known to reflect network  topology  \cite{belkin2001laplacian,chung1997spectral,peixoto2013eigenvalue,taylor2016synchronization}. 
In particular, previous research studying a random-network ensemble  found VNE to be larger for degree-regular networks and smaller for networks with irregular structures such as long paths and nontrivial symmetries \cite{passerini2008neumann}.  VNE has been receiving growing attention for its utility for network-network comparison and has been used recently to  hierarchically cluster layers in multilayer networks  \cite{de2016spectral,de2015structural}.

\begin{figure}
\centering
\includegraphics[width=\linewidth]{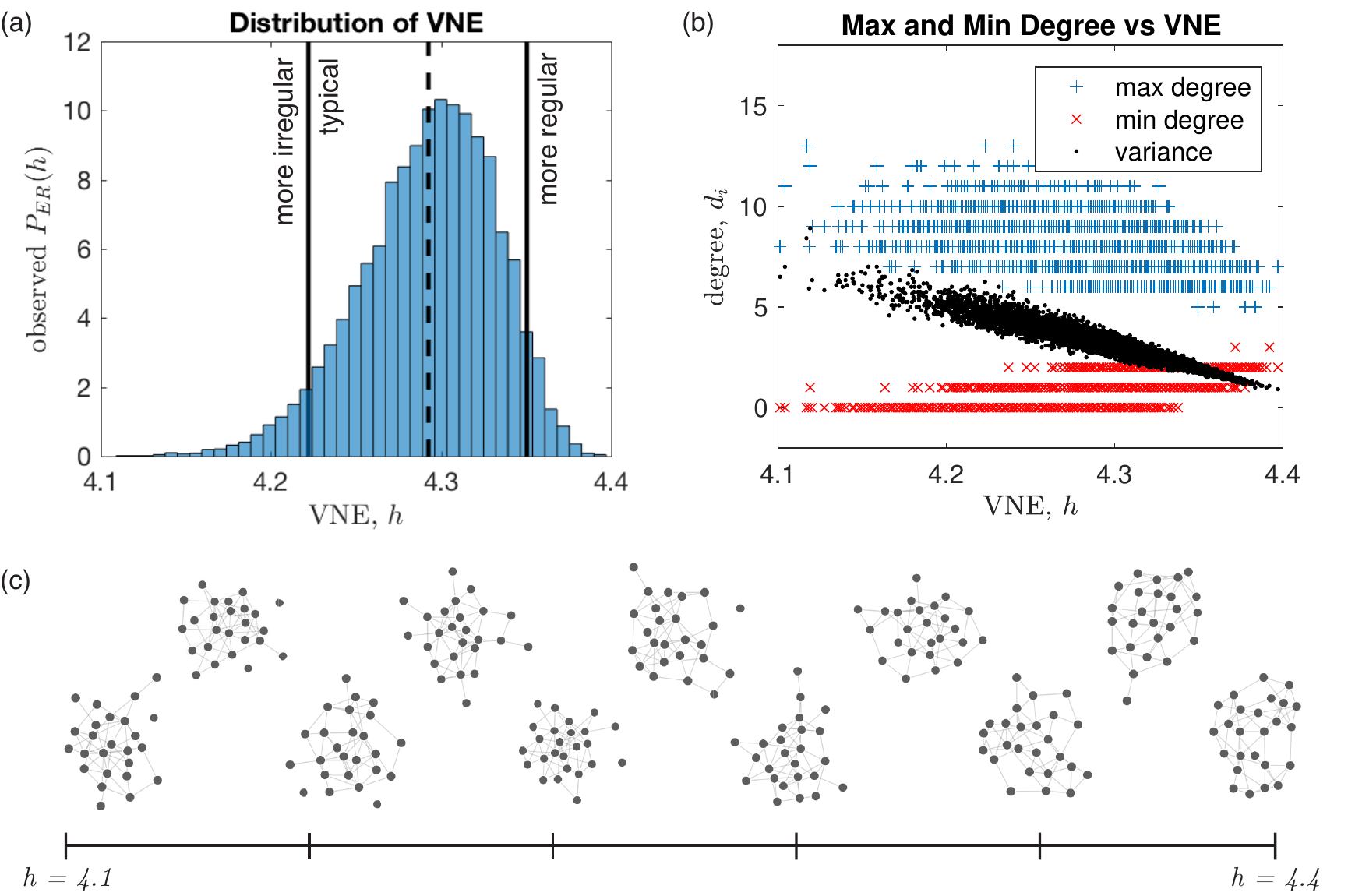}
\caption{Network-ensemble comparison with von Neumann Entropy (VNE) distinguishes typical and atypical networks according to their  `irregularity\drt{'.} 
(a)~Empirical distribution $P^{(N,M)}(h)$ of VNE for $10^4$ networks sampled from an Erd\H{o}s-R\'enyi (ER) random-network ensemble $\mathbf{G}_{N,M}$ with $N=25$ nodes and $M=50$ undirected edges.
(b)~Degree heterogeneity negatively correlates with VNE.
(c)~Example networks sampled from $\mathbf{G}_{N,M}$ are arranged according to VNE. 
}
\label{fig:vne1}
\end{figure}

As motivation, we present a numerical experiment illustrating the ability of VNE to distinguish between typical and atypical networks in the Erd\H{o}s-R\'enyi $\mathbf{G}_{N,M}$ ensemble of simple random networks (see Definition~\ref{def:ER} in Sec.~\ref{sec:Ensemble}) with $N=25$ and $M=50$.  We studied the probability distribution of VNE across the ensemble, $P^{(N,M)}(h)$,  which we estimated by sampling $10^4$ networks from the ensemble. In Fig.~\ref{fig:vne1}(a), we plot the observed distribution $P^{(N,M)}(h)$. The vertical dashed line indicates the empirical mean and solid lines indicate the 5\% and 95\%  quantiles. In Fig.~\ref{fig:vne1}(b), we provide a scatter plot that indicates for each of these networks the maximum degree, minimum degree, and degree variance versus VNE. Note that degree heterogeneity negatively correlates with VNE, illustrating that networks with small (large) VNE are more (less) degree irregular.  In Fig.~\ref{fig:vne1}(c), we provide visualizations for several networks sampled from $\mathbf{G}_{N,M}$, which are arranged so that their VNEs increase from left to right.

The main takeaway from this experiment is that larger VNE can be interpreted to indicate decreased `irregularity' in a network. By studying the distribution of VNE for a random-network ensemble, one can differentiate between networks that are typical (i.e., their VNEs are typical and they contain a typical amount of irregular structure) and those that are atypical (i.e., their VNEs are in the tails of the distribution $P^{(N,M)}(h)$, with either more or less irregularity than is typical).  It is worth noting that although this experiment focuses on degree heterogeneity, previous research has established a complicated connection between Laplacian spectra (and hence VNE) and other sources for structural irregularity including subgraph motifs \cite{preciado2013moment,preciado2015moment}, communities \cite{fiedler1989laplacian,peixoto2013eigenvalue}, manifolds \cite{belkin2001laplacian,coifman2006harmonic}, trees/loops  \cite{nakatsukasa2013mysteries,saito2011phase,schultz2016tweaking,taylor2016synchronization}, and so on.
\drt{See \cite{anand2011shannon,anand2009entropy} for connections between VNE and other information-theoretic measures of networks.}

\clearpage

\subsection{Random-Network Ensembles}\label{sec:Ensemble}

We consider two random-network ensembles that have each received considerable attention: the ER and configuration model ensembles.

\begin{definition}[Erd\H{o}s-R\'enyi Ensemble $\mathbf{G}_{N,M}$ of Simple Networks  \cite{erdos1959random}]\label{def:ER}
Let $\mathcal{G} _{N,M} = \{ {G}_s\}$ denote the set of networks $G_s$ with $N$ nodes and $M$ undirected edges, disallowing repeat edges and self-edges. Note that $S_{N,M}=| \{ \mathcal{G}_{N,M}\}| =\binom{N(N-1)/2}{M}$. Let ${\bf \pi}$ denote a uniform distribution on $\mathcal{G}_{N,M}$ so that $\pi_s=1/S_{N,M}$. The ER random-network ensemble is defined by the pair $\mathbf{G}_{N,M}=(\mathcal{G}_{N,M},{\bf \pi})$.
\end{definition}

The most common approach to sample networks from $\mathbf{G}_{N,M}$ involves enumerating the potential edges $\{1,2,\dots,\binom{N}{2}\}$ and choosing $M$ of them uniformly at random. Whenever considering ER models, it is typically worth noting that there exists another ER model that is closely related, $\mathbf{G}_{N,p}$, in which each pair of nodes is connected by an edge independently with probability $p\in[0,1]$. $\mathbf{G}_{N,M}$ and $\mathbf{G}_{N,p}$ are  referred to as the \emph{microcanonical} and \emph{canonical} ER models, respectively. They have greatly benefited theory development in network science and graph theory and are arguably the simplest random-network ensembles. Real-world networks, however, are well known to contain a variety of structures  not well represented by the ER ensembles. In particular,  the probability distribution of node degrees is binomial for the ER models; however, the degree distributions of empirical networks are often much more heavy-tailed (networks with power-law distributions \cite{barabasi1999emergence} being just one such class). For this reason, there is widespread interest in configuration model random-network ensembles that allow {\it a priori} specification of the degree sequence $\mathbf{d}=\{d_i\}$. 

\begin{definition}[Configuration Ensemble $\drt{\hat{\mathbf{G}}_{N,\mathbf{d}}}$ of Simple Networks \cite{bekessy1972asymptotic}]\label{def:config}
Let $\drt{\hat{\mathcal{G}}}_{N,\mathbf{d}} = \{ \hat{G}_s\}$ denote the set of networks $\hat{G}_s$ with $N$ nodes and degree sequence $\mathbf{d}$, disallowing  repeat edges and self-edges. Note that $M=\frac{1}{2}\sum_i d_i$. 
Let $\hat{ \pi}$ denote a uniform distribution on $\drt{\hat{\mathcal{G}}}_{N,\mathbf{d}}$  with $ \hat{\pi}_s=\drt{ \hat{S}}_{N,\mathbf{d}}^{-1}$ and $\drt{ \hat{S}}_{N,\mathbf{d}}=|\drt{\hat{\mathcal{G}}}_{N,\mathbf{d}}|$.
The configuration ensemble of simple random networks is defined by the pair $\drt{\hat{\mathbf{G}}}_{N,\mathbf{d}}=(\drt{\hat{\mathcal{G}}}_{N,\mathbf{d}},\hat{\pi})$.
\drt{For further clarity, we use hats on not-already-accented mathematical objects associated with the configuration ensemble to clearly delineate those objects from analogous objects for the ER ensemble.}
\end{definition}

\begin{figure}[t]
\centering
\includegraphics[width=.5\linewidth]{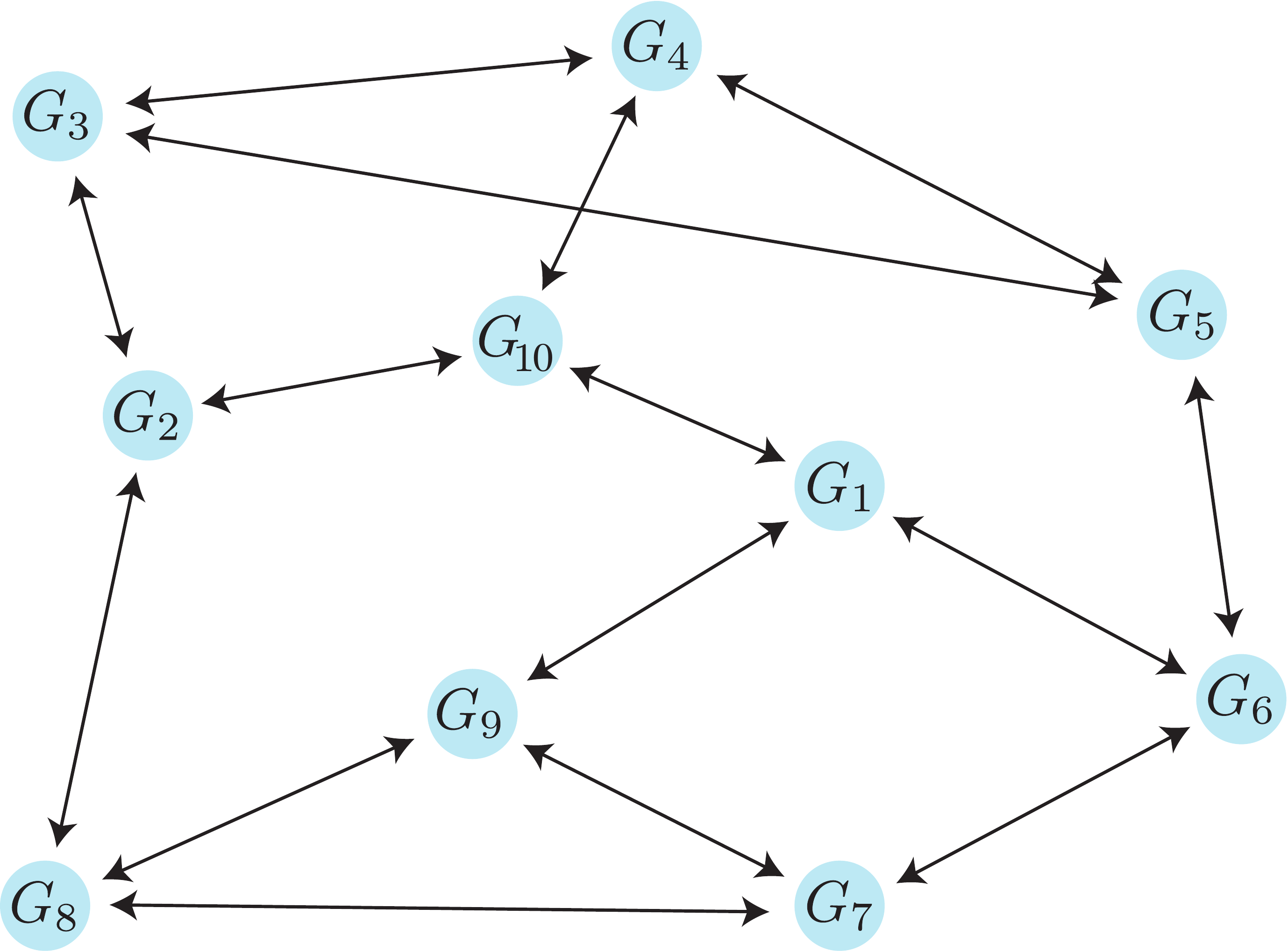}
\caption{
Networks $\{G^{(t)}\}$ obtained after $t$ stochastic rewires of an original network $G^{(0)}$ are modeled by a random walk on a set of networks $\{{G}_s\}$. Letting $\mathbf{\pi}^{(t)}_s $ denote the probability that $G^{(t)}=G_s$ and $P^{(t)}(h) $ denote the probability that $h(G^{(t)})=h$, we will study the evolution of $\mathbf{\pi}^{(t)}$ and $P^{(t)}(h)$ using Markov chains.
}
\label{fig:chain}
\end{figure}

There are two main classes of algorithms for sampling random networks from $\drt{\hat{\mathcal{G}}}_{N,\mathbf{d}} $: random-matching methods and Markov chain Monte Carlo (MCMC) methods  \cite{blitzstein2011sequential,milo2003uniform}. The random-matching methods involve enumerating the $d_i$ ``stubs'' of edges for each node $i$ and then randomly matching pairs of stubs of different nodes in a ``configuration'' of allowable network edges \cite{bekessy1972asymptotic,bollobas1980probabilistic,steger1999generating}. We note that only some degree sequences $\mathbf{d}$ are \emph{graphical} in that there exist graphs for such a degree sequence \cite{arratia2005likely}. In contrast, the MCMC methods involve taking an initial network and randomizing it via repeated stochastic rewiring, which can be studied as a Markov chain in which states represent networks and transitions represent rewiring \cite{cooper2005sampling,jerrum1990fast} (see Fig.~\ref{fig:chain}). In general, there are many choices for how to implement stochastic rewiring, which can give rise to various random-network ensembles. We describe in the next section an MCMC rewiring process that converges in the limit of many rewires to uniform sampling \cite{fosdick2016configuring} (consistent with Definition \ref{def:config}).

Finally, we note that there exist many other generative models for constructing random networks---stochastic block models \cite{stanley2015clustering,peixoto2013eigenvalue}, exponential random graphs \cite{kolaczyk2014statistical}, and so on (see reviews \cite{boccaletti2014structure,holme2012temporal,newman2010networks} and references therein)---that introduce different constraints  aimed toward diverse applications.

\subsection{Degree-Preserved Rewiring as a Markov Chain on a Set of Networks}\label{sec:Markov}

We now describe a stochastic rewiring process called \emph{double-edge-swap vertex-labeled rewiring} \cite{fosdick2016configuring} that can be used for MCMC sampling of the configuration ensemble given by Definition \ref{def:config}. Herein, we refer to the process as degree-preserved rewiring.
\drt{
We  note  that this rewiring process has been unknowingly reinvented several times by different researchers from different fields, and this work is reviewed and expanded on in  \cite{fosdick2016configuring}. }

~\\

\begin{definition}[Degree-Preserved Rewiring \cite{fosdick2016configuring}]\label{def:DP_rewire} \drt{D}egree-preserved rewiring is a stochastic map $T_{DP}: \drt{\hat{\mathcal{G}}}_{N,\mathbf{d}} \to \drt{\hat{\mathcal{G}}}_{N,\mathbf{d}}$ defined by  $(\mathcal{V},\mathcal{E})\to(\mathcal{V},\mathcal{E}')$, where $\mathcal{E}'$ is given by the following stochastic process. Choose two unique edges $(i,j)$ and $ (i',j')$ uniformly at random from $\mathcal{E}$\drt{.} Consider a proposed edge swap in which two edges $\mathcal{E}^{(-)}= \{(i,j),(i',j')\}$ are removed and two new edges $\mathcal{E}^{(+)}$ are added, uniformly at random selecting between $\mathcal{E}^{(+)}=\{(i,j'),(i',j)\}$ and $\mathcal{E}^{(+)}=\{(i,i'),(j',j)\}$. If the proposed  new edges $\mathcal{E}^{(+)}$ do not give rise to a self edge nor a repeat edge (which are disallowed by $\drt{\hat{\mathcal{G}}}_{N,\mathbf{d}}$),  then the proposed edge swap is implemented,  $\mathcal{E}\mapsto (\mathcal{E}\setminus\mathcal{E}^{(-)})\cup\mathcal{E}^{(+)}$. Otherwise, the network is left unchanged, $\mathcal{E}\mapsto \mathcal{E}$.
\end{definition}

Because $N$ and $\mathbf{d}$ are invariant under degree-preserved rewiring, iterative rewiring can be modeled as a random walk on the set $\drt{\hat{\mathcal{G}}}_{N,\mathbf{d}}=\{\hat{G}_s\}$, which we enumerate using $s\in\{1,\dots,\drt{\hat{S}}_{N,\mathbf{d}}\}$. We let  $\hat{ {G}}^{(0)}\in \drt{\hat{\mathcal{G}}}_{N,\mathbf{d}}$ denote an original (e.g., empirical) network, $\drt{\hat{G}}^{(t)}\in\drt{\hat{\mathcal{G}}}_{N,\mathbf{d}}$ denote a network after $t$ rewires,  $\hat{\pi}_s^{(t)}$ denote the probability that $\hat{ {G}}^{(t)}=\hat{ {G}}_s$. Note that ${\bf \hat{\pi}}^{(0)}_s=1$ for $s$ such that $\drt{\hat{G}}^{(0)} = \drt{\hat{G}}_s$ and ${\bf \hat{\pi}}^{(0)}_s=0$ otherwise. The evolution of ${\bf \hat{\pi}}^{(t)}$ is given by
\begin{equation}
\hat{\pi}_s^{(t+1)} = \sum_r \hat{\pi}_r^{(t)} \hat{P}_{rs}, \label{eq:DP_chain}
\end{equation}
where  $\hat{P}_{rs}$ is a transition matrix describing the probability that network $\drt{\hat{G}}_r$ will become $\drt{\hat{G}}_s$ after a degree-preserved rewire.

\begin{theorem}[Markov Chain Convergence for Degree-Preserved Rewiring \cite{fosdick2016configuring}] \label{thm:bailey}
The Markov chain defined by Eq.~\ref{eq:DP_chain} is ergodic and has uniform stationary distribution 
\begin{equation}
\lim_{t\to\infty} \hat{{ \pi}}_s^{(t)} = \drt{\hat{{S}}}_{N,\mathbf{d}}^{-1}.
\end{equation} 
\begin{proof}
The result follows from showing that the Markov chain is connected, aperiodic and degree regular  \cite{fosdick2016configuring}.
\end{proof}
\end{theorem}
\begin{remark} 
\drt{See \cite{greenhill2014switch,greenhill2017switch} for research relating the rate of convergence to properties of the  set  of degrees  that is preserved under rewiring.}
\end{remark}

\section{Main Results}\label{sec:Main}
\drt{Our approach for efficient network-ensemble  comparison involves using a perturbation analysis of stochastic rewiring processes associated with random-network ensembles.}
In Sec.~\ref{sec:NDP}, we define and analyze a uniform stochastic-rewiring process.
In Sec.~\ref{sec:CMT}, we analyze the distributional convergence of network summary statistics (including VNE) for networks undergoing stochastic rewiring.
In Sec.~\ref{sec:Cont}, we develop a first-order perturbation analysis for the effect of rewiring on VNE.
In Sec.~\ref{sec:Expected}, we study the expected perturbation to VNE and the Laplacian matrix under uniform rewiring.
\drt{In Sec.~\ref{sec:Roadmap}, we provide a roadmap for how this approach can be extended to other rewiring processes and other summary statistics.}

\subsection{ Convergence of Network Ensembles obtained under Stochastic  Rewiring}\label{sec:NDP}
In this section, we define and study convergence for a sequences $\{\mathbf{G}^{(t)}\}$ of random-network ensembles arising from two stochastic rewiring processes. We begin by defining another stochastic rewiring process: uniform  rewiring.

\begin{definition}[Uniform Rewiring]\label{def:NDP_rewire}
Uniform rewiring is a stochastic map $T_{U}: \mathcal{G}_{N,M} \to \mathcal{G}_{N,M}$ defined by   $(\mathcal{V},\mathcal{E})\to(\mathcal{V},\mathcal{E}')$, where $\mathcal{E}'$ is given by the following stochastic process. Choose uniformly at random an edge $(i,j)\in\mathcal{E}$ and remove it from $\mathcal{E}$. Then choose uniformly at random a new edge $(i',j')$ from the $N(N-1)/2-M +1$ possible edges outside of $\mathcal{E}\setminus(i,j)$ [that is, allowing re-selection of $(i,j)$], and add the edge to $\mathcal{E}$. It follows that $\mathcal{E}\mapsto (\mathcal{E}\setminus (i,j))\cup (i',j')$.
\end{definition}

Because $N$ and $M$ are invariant under uniform rewiring, iterative uniform rewiring can be modeled as a random walk on the set $\mathcal{G}_{N,M}=\{ {G}_s\}$, which we enumerate using $s\in\{1,\dots,S_{N,M}\}$. We let  $G^{(t)} \in\mathcal{G}_{N,M}$ denote a network after it undergoes $t$ rewires and ${\pi}_s^{(t)}$ denote the probability that $ \drt{G}^{(t)}= \drt{G}_s$. Obviously, ${\bf{\pi}}^{(0)}_s=1$ for $s$ such that $ {\drt{G}}^{(0)}= {\drt{G}}_s$ and ${\bf  {\pi}}^{(0)}_s=0$ otherwise.  The evolution of ${\bf {\pi}}^{(t)}$ is given by
\begin{equation}
 {\pi}_s^{(t+1)} = \sum_r  {\pi}_r^{(t)}  {P}_{rs}, \label{eq:NDP_chain}
\end{equation}
where  ${P}_{rs}$ is a transition matrix describing the probability that network $G_r$ will become $G_s$ after a uniform rewire.

We identify the following limiting behavior for ${\bf \pi}^{(t)}$.

\begin{theorem}[Convergence of Uniform Rewiring] \label{thm:CONVERGENCE}
The Markov chain for uniform rewiring \eqref{eq:NDP_chain} is ergodic and has uniform stationary distribution 
\begin{equation}
\lim_{t\to\infty} { \pi}_s^{(t)} = S_{N,M}^{-1}. \label{eq:NDP_limit}
\end{equation} 
\begin{proof}
See Appendix \ref{appendixA0}.
\end{proof}
\end{theorem}

We now define a notion of convergence for a sequence $\{\mathbf{G}^{(t)}\}$ of random-network ensembles.

\begin{definition}[Convergence of Random-Network Ensembles]\label{def:CONV}
Let $\{\mathbf{G}^{(t)}\}$ denote a sequence of random-network ensembles in which  $\mathbf{G}^{(t)} = (\{G_s\},\pi^{(t)})$. We say that $\mathbf{G}^{(t)}$ converges to $\mathbf{G} = (\{G_s\},\pi)$ iff 
\begin{equation}
 \mathbf{G}^{(t)} \to \mathbf{G}  \Leftrightarrow \pi^{(t)}\to\pi .
\end{equation}
\end{definition}

We are now ready to describe the convergence of random-network ensembles arising from stochastic uniform and degree-preserved rewiring.  
\begin{corollary}[ER Ensemble Convergence]\label{corr:NDP_conv}
Consider the ensemble of random networks $\mathbf{G}^{(t)} =(\mathcal{G}_{N,M},\pi^{(t)})$ obtained by $t$ uniform rewires of an initial network $G^{(0)}\in\mathcal{G}_{N,M}$. The sequence $\{\mathbf{G}^{(t)} \}$ of ensembles converges to the ER ensemble given by Definition \ref{def:ER},
\begin{equation}
\lim_{t\to\infty} \mathbf{G}^{(t)} = \mathbf{G}_{N,M}.
\end{equation}
\begin{proof}
The result follows straightforwardly from Theorem~\ref{thm:CONVERGENCE}.
\end{proof}
\end{corollary}

\begin{corollary}[Configuration-Model Ensemble Convergence]\label{corr:DP_conv}
Consider the ensemble of random networks $\hat{\mathbf{G}}^{(t)} =( \drt{\hat{{\mathcal{G}}}}_{N,\mathbf{d}},\hat{\pi}^{(t)})$ that is obtained by $t$ degree-preserved rewires of an initial network $\hat{G}^{(0)} \in \drt{\hat{\mathcal{G}}}_{N,\mathbf{d}}$.  The sequence $\{\hat{\mathbf{G}}^{(t)} \}$, \drt{where $\hat{\mathbf{G}}^{(t)} = (\{\hat{G}_s\},\drt{\hat{\pi}}^{(t)})$,} converges to the configuration model ensemble given by Definition \ref{def:config},
\begin{equation}
\lim_{t\to\infty} \hat{\mathbf{G}}^{(t)} =  \drt{\hat{{\mathbf{G}}}}_{N,\mathbf{d}}.
\end{equation}
\begin{proof}
The result follows straightforwardly from Theorem~\ref{thm:bailey}.
\end{proof}
\end{corollary}

\subsection{Distributional Convergence of Network Summary Statistics}\label{sec:CMT}

We now study the distribution of VNE and other  summary statistics for random-network ensembles associated with uniform and degree-preserved rewiring.

\begin{theorem}[Distributional Convergence of Network Statistics] \label{thm:afff}
Let $\{\pi^{(t)}\}$ and $\{{\hat{\pi}}^{(t)}\}$ describe sequences of probability distributions over $\mathcal{G}_{N,M}$ and $\drt{\hat{\mathcal{G}}}_{N,\mathbf{d}}$, respectively, for the uniform and degree-preserved rewiring processes. Further, let $f: \{G_s\} \mapsto \mathbb{R}$  denote any scalar-valued function on a network and let 
\begin{align}
P^{(t)}(f) &= 
\sum_{s=1}^{S_{N,M}} \pi_s^{(t)}\delta_{f(G_s)}(f) \nonumber\\
{\hat{P}}^{(t)}(f) &= 
 \sum_{s=1}^{\drt{\hat{S}}_{N,\mathbf{d}}} {\hat{\pi}}_s^{(t)} \delta_{f(\drt{\hat{G}}_s)}(f) \label{eq:limt}
\end{align}
denote the respective distributions of $f$ across the associated random-network ensembles $\mathbf{G}^{(t)}$ and  $\hat{\mathbf{G}}^{(t)}$. Here,  $\delta_{g}(f)$ is the Dirac delta function with weight concentrated at $f=g$ [i.e., $\delta_g(f)=\delta(f-g)$]. The following limits converge in distribution as $t\to\infty$ 
\begin{align}\label{eq:llim}
P^{(t)}(f) &\overset{d}{\to} P^{(N,M)}(f) \nonumber \\
\hat{P}^{(t)}(f) &\overset{d}{\to} \drt{\hat{P}}^{(N,\mathbf{d})}(f),
\end{align}
where $P^{(N,M)}(f)$ and $\drt{\hat{P}}^{(N,\mathbf{d})}(f)$ denote the distributions of $f(G)$ for the ER and configuration random-network ensembles, respectively.
\begin{proof}
We take the limit of both sides of the equations in \eqref{eq:limt}. Because the summations are finite, the limits can be taken inside the summation. The equations in \eqref{eq:llim} follow directly from Eqs.~\eqref{corr:NDP_conv} and \eqref{corr:DP_conv}. 
\end{proof}
\end{theorem}

\begin{corollary}[Distributional Convergence of VNE] \label{cor:14d}
Letting $f: \{G_s\} \mapsto \mathbb{R}$ denote VNE, $h(G_s)$, given by Definition~\ref{def:VNE}, Eq.~\eqref{eq:llim} implies
\begin{align}\label{eq:llim2}
P^{(t)}(h) &\overset{d}{\to} P^{(N,M)}(h) \nonumber\\
\hat{P}^{(t)}(h) &\overset{d}{\to} \drt{\hat{P}}^{(N,\mathbf{d})}(h),
\end{align}
where $P^{(N,M)}(h)$ and $\drt{\hat{{P}}}^{(N,\mathbf{d})}(h)$ denote the distributions of VNE for the ER and configuration random-network ensembles, respectively.

\end{corollary}

\subsection{Perturbation of VNE and Laplacian Matrices}\label{sec:Cont}
Having characterized the long-time behavior (i.e., after many rewires)  of networks subjected to uniform and degree-preserving  rewiring processes (as well as their associated network statistics such as VNE), we now turn our attention to studying the effect on VNE due to a small number of rewires. To this end,  in this section we develop a first-order perturbation analysis for VNE. We begin by presenting a well-known result that describes the first-order perturbation of eigenvalues and eigenvectors of a symmetric matrix, which we present for an unnormalized Laplacian $ {L}$.

\begin{theorem}[Perturbation of Simple Eigenvalues and their Eigenvectors \cite{atkinson2008introduction}] 
Let $ {L}$ be a symmetric $N\times N$ matrix with simple eigenvalues $\{\lambda_i\}$ and normalized eigenvectors $\{\bm v^{(i)}\}$.
Consider a fixed symmetric perturbation matrix $\Delta  {L}$, and let $ {L}(\epsilon)= {L}+\epsilon \Delta  {L}$.
We denote the eigenvalues  {and eigenvectors} of $ {L}(\epsilon)$ by $\lambda_n(\epsilon)$  {and $\bm{v}^{(n)}(\epsilon)$, respectively, for $i=1,2,\dots,N$.
It follows that
\begin{align}
\lambda_i(\epsilon)  &= \lambda_i + \epsilon \lambda'_i(0) + \mathcal{O}(\epsilon^2),\nonumber \\
{\bm v^{(i)}} (\epsilon) &= \bm v^{(i)} + \epsilon {\bm{v}^{(i)}}'(0) + \mathcal{O}(\epsilon^2),
\label{eq:eigen_perturb_a}
\end{align}
and the derivatives with respect to $\epsilon$ at $\epsilon=0$ are given by
\begin{align}
\lambda_i'(0) & = (\bm v^{(i)})^T \Delta  {L} \bm v^{(i)}\nonumber \\
 {\bm{v}^{(i)}}'(0) &= \sum_{{j\not=i}} \frac{ (\bm v^{(j)})^T \Delta  {L} \bm v^{(i)}}{\lambda_i-\lambda_j} \bm{v}^{(j)}.
\label{eq:eigen_perturb_b}
\end{align}}
\end{theorem}
\begin{proof}
See \cite{atkinson2008introduction}.
\end{proof}
\begin{remark} 
For the unnormalized Laplacian matrix $L$, $\lambda_1(\epsilon)=0$ and $\bm{v}^{(1)}(\epsilon)=N^{1/2}\bm{1}$ for all values of $\epsilon$. Any allowable perturbation matrix $\Delta  {L}$ will have the same null space as $ {L}$, $\text{span}(\bm{1}) $, and so  $\lambda_1'(0)=0$ and ${\bm{v}'}^{(1)}(0)=\bm{0}$.
\end{remark} 
\begin{remark} 
The first-order approximations in Eq.~\eqref{eq:eigen_perturb_a} are accurate when the perturbations are small. However, the regime for which this approximation is valid (i.e., how small $\epsilon$ needs to be) generally depends on $ {L}$, $\epsilon$, and the perturbation  $\Delta  {L}$. Accuracy typically requires $\epsilon\lambda'_i(0)/\lambda_i$ to be small \cite{taylor2016synchronization}. 
\end{remark} 

We now present a first-order perturbation analysis of the VNE for a network subjected to a modification.
\begin{theorem}[First-Order Perturbation of VNE] \label{thm:main1}
Let $h(0)$ denote the VNE given by Definition \ref{def:VNE} for an unnormalized network Laplacian ${L}$ with simple eigenvalues $\{\lambda_i\}$, and let $h(\epsilon)$ denote the VNE for the network after it undergoes a network modification encoded by $ {L}(\epsilon) =  {L} + \epsilon \Delta  {L}$. We assume the eigenvalues of $ {L}(\epsilon)$ are simple.
The first-order expansion in $\epsilon$ for the perturbed VNE is 
\begin{equation} \label{eq:first_VNE}
h( \epsilon) = h(0) + \epsilon h'(0) + \mathcal{O}(\epsilon^2)\drt{,}
\end{equation}
where
\begin{equation} \label{eq:vne_perturb}
h'(0) = -\frac{1}{2M} \sum_{i }   (\bm v^{(i)})^T \Delta  {L} \bm v^{(i)}  \left( \log_2 \left( \frac{\lambda_i}{2M} \right) + \frac{1}{\ln(2)} \right) .
\end{equation}
\begin{proof}
See Appendix \ref{appendixA}
\end{proof}
\end{theorem}

\begin{remark}
The Laplacian matrix $L$ for networks consisting of $k$ connected components will have $K$ eigenvalues $\lambda_1=\lambda_2=\dots=\lambda_k=0$. In this case, as well as other scenarios with eigenvalues having multiplicity greater than or equal to two, Eqs.~\eqref{eq:eigen_perturb_a}--\eqref{eq:vne_perturb} can be used to estimate the perturbation of the remaining simple eigenvalues that have multiplicity one, and for which $\frac{1}{\lambda_i-\lambda_j}$ is guaranteed to be finite.
\end{remark}

\begin{corollary}[Edge Perturbation of von Neumann Entropy] \label{cor:1edge}
When the network is modified by adding (+) or removing (-) an unweighted edge $(p,q)$, the Laplacian perturbation matrix takes the form
\begin{align}
\Delta  {L}_{ij}^{(pq)} = \left\{ \begin{array}{rl} 
\pm 1 ,&  {(i,j)\in\{(p,p),(q,q)\}}\\ 
\mp 1 ,&  {(i,j)\in\{(p,q),(q,p)\}}\\ 
0,&\text{otherwise}, 
\end{array}\right. 
\label{eq:DL}
\end{align}
and
Eq.~\eqref{eq:vne_perturb} can be simplified as  
\begin{equation} \label{eq:vne_perturb2}
h'(0) = - \frac{1}{2M} \sum_{i= 1}^{N}     \pm (\bm v^{(i)}_p-\bm v^{(i)}_q)^2 
\left( \log_2 \left( \frac{\lambda_i}{2M} \right) + \frac{1}{\ln(2)} \right)    ,
\end{equation}
where $\pm$ corresponds to addition and removal, respectively.
 \begin{proof}
 See Appendix \ref{appendixB}
\end{proof}
\end{corollary}

\begin{corollary}[Edge-Set Perturbation of von Neumann Entropy] \label{cor:edgeset}
When the network is modified by adding a set of edges $\mathcal{E}^+$ and removing a set $\mathcal{E}^-$, the Laplacian perturbation matrix takes the form
\begin{align}
\Delta  {L}_{ij}^{(\mathcal{E}^+,\mathcal{E}^-)} = \sum_{(p,q)\in\mathcal{E}^+}  \Delta  {L}_{ij}^{(pq)} - \sum_{(p,q)\in\mathcal{E}^-}  \Delta  {L}_{ij}^{(pq)},
\label{eq:DL_2}
\end{align}
where $ \Delta  {L}_{ij}^{(pq)}$ is given by Eq.~\eqref{eq:DL}, and Eq.~\eqref{eq:vne_perturb} becomes  
\begin{eqnarray} \label{eq:vne_perturb3}
h'(0) &=& - \frac{1}{2M} \sum_{i = 1}^{N}   
\left(\sum_{(p,q)\in\mathcal{E}^+  }   (\bm v^{(i)}_p-\bm v^{(i)}_q)^2  - \sum_{(p,q)\in\mathcal{E}^-  }   (\bm v^{(i)}_p-\bm v^{(i)}_q)^2 \right)     \nonumber\\
&~&  \hspace*{2in} \times \left( \log_2 \left( \frac{\lambda_i}{2M} \right) + \frac{1}{\ln(2)} \right).
\end{eqnarray}
 \begin{proof}
The proof is straightforward using the linearity property of edge additions and removals.
\end{proof}
\end{corollary}

\subsection{Expected Perturbations under Uniform Rewiring}\label{sec:Expected}

In this section, we describe the expected perturbations due to uniform rewiring. We note that solving the expected perturbations under degree-preserved stochastic rewiring is much more difficult and is left for future research.

\begin{theorem}[Expected Change to the Laplacian under Uniform Rewiring] \label{thm:expected}
Consider an undirected unweighted network $\drt{G}$ with $N$ nodes, $M$ edges, 
adjacency matrix $A$, node degrees $\{d_i\}$, and
Laplacian matrix $L$. The expected change $\Delta L$ to $L$ under uniform rewiring (see Definition \ref{def:NDP_rewire}) is given by
\begin{equation} \label{eq:expected_laplacian}
\mathbf{E} [ \Delta L_{ij}] = 
\begin{cases}  \frac{N - 1- d_i}{\frac{N(N-1)}{2} - M +1} - \frac{d_i}{M} &\mbox{if } i = j\\
\frac{A_{ij}}{M} - \frac{1-A_{ij}}{\frac{N(N-1)}{2} - M+1} &\mbox{if } i \neq j
\end{cases} 
\end{equation}
\begin{proof}
See Appendix \ref{appendixC}
\end{proof}
\end{theorem}

\begin{corollary}[Expected First-Order Perturbation under Uniform Rewiring] \label{thm:first}
Under uniform rewiring (see Definition \ref{def:NDP_rewire}), the expected first-order terms for $\lambda_i(\epsilon)$ and ${\bm v^{(i)}} (\epsilon)$ [see Eq.~\eqref{eq:eigen_perturb_b}] and $h(\epsilon)$ [see Eq.~\eqref{eq:vne_perturb}] are given by
\begin{align}
\mathbf{E} [\lambda_n'(0)] & = (\bm v^{(n)})^T \mathbf{E} [\Delta L] \bm v^{(n)}  \\ 
\mathbf{E} [{\bm{v}^{(n)}}'(0)] &= \sum_{{m\not=n}} \frac{ (\bm v^{(m)})^T \mathbf{E} [\Delta L] \bm v^{(n)}}{\lambda_n-\lambda_m} \bm{v}^{(m)} \\
\mathbf{E} [h'(0)] &= - \sum_{n = 1}^{N} \left( \frac{ (\bm v^{(n)})^T \mathbf{E} [\Delta L] \bm v^{(n)}}{2M} \right) \left( \log_2 \left( \frac{\lambda_n}{2M} \right) + \frac{1}{\ln(2)} \right) ,
\label{eq:expected_eigen_perturb}
\end{align}
where $\mathbf{E} [\Delta L]$ is given by Eq.~\eqref{eq:expected_laplacian}.
\begin{proof}
We take the expectation of Eqs.~\eqref{eq:eigen_perturb_b} and \eqref{eq:vne_perturb}, use the linearity property for expectation, and combine these results with Eq.~\eqref{eq:expected_laplacian}.
\end{proof}
\end{corollary}

\subsection{\drt{Road Map for Other Rewiring Processes and Other Summary Statistics}}\label{sec:Roadmap}

\drt{
The analytical approximations given by Thm.~\ref{thm:expected} and Corr.~\ref{thm:first} could be obtained  due to the simplicity of  uniform rewiring (see Definition~\ref{def:NDP_rewire}). 
%
However, this is not the case for degree-preserved rewiring (see Definition~\ref{def:DP_rewire}), for which analytical predictions for $\mathbf{E} [ \Delta L_{ij}] $ and ${\bf E}[h'(0]$ are beyond the scope of this paper. A main difficulty stems from the observation that the probability of rewiring a given edge $(a,b)$ depends on whether or not it is swapped with another edge $(c,d)$. Given a pair of edges, $\{ (a,b),(c,d)\}$, the probability of swapping $\{ (a,b),(c,d)\} \to \{ (a,d),(b,c)\}$ or $\{ (a,b),(c,d)\} \to \{ (a,c),(b,d)\}$, or not swapping at all, depends on whether or not the edges $(a,c)$, $(a,d)$, $(b,c)$, and $(b,d)$ already exist. 
Moreover, for a given edge $(a,b)$, the expected change $\mathbf{E} [ \Delta L_{ab}] $ requires considering the probability of swapping with all other edges $\{(c,d)\}$, and this pursuit quickly becomes intractable for large networks.
Developing perturbation theory for degree-preserved rewiring (and other stochastic rewiring processes \cite{fosdick2016configuring}) remains an important direction for future research.

Nevertheless, by iteratively simulating one single rewire, one can still numerically estimate ${\bf E}[h'(0)]$ (or the expected change to other summary statistics) for degree-preserved rewiring  and other rewiring processes.
In the following section, we will make use of these analytical and numerical predictions for ${\bf E}[h'(0)]$ to estimate the number of rewires necessary for empirical networks to resemble typical networks from  ER and configuration-model network ensembles.

}

\section{Numerical Experiments}\label{sec:Num}
We now present numerical experiments supporting and demonstrating the utility of our results from Sec.~\ref{sec:Main}.
In Sec.~\ref{sec:num_pert}, we support our perturbation results describing how network modifications affect VNE.
In Sec.~\ref{sec:num_conv}, we support our results for the distributional convergence of VNE for stochastic uniform and degree-preserved rewiring processes.
In Sec.~\ref{sec:num_empirical}, we highlight an application of our analysis: network-ensemble comparison for empirical networks.

\begin{figure}
\centering
\includegraphics[width=.9\linewidth]{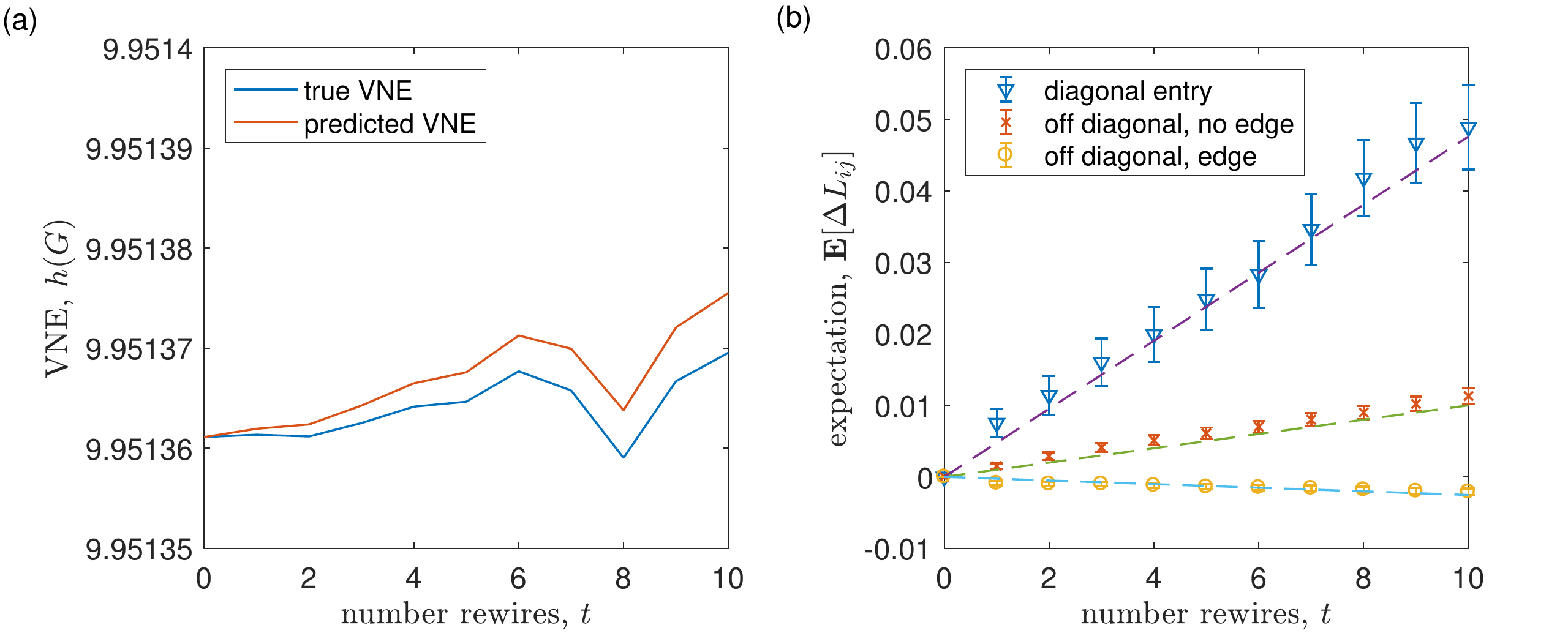}
\caption{
(a)  True values of VNE, $h(G^{(t)})$, for a network subjected to $t$ uniform rewires compared to the first-order approximation given by Theorem \ref{thm:main1}, which uses the known perturbation of the Laplacian matrix, $\Delta L$.
(b)  Comparison of the observed mean ${\mathbf{E}}[\Delta L_{ij}^{(t)}]$ (symbols) to its expectation  given by Theorem \ref{thm:expected} (dashed lines) for three entries: a diagonal entry $\Delta L_{ii}$ and two entries that correspond to the absence and presence, respectively, of an edge $(i,j)$ in network $\drt{G^{(0)}}$. Error bars indicate standard error across $K=10,000$ trials of uniform rewiring.
}
\label{fig:VNE_perturb_1}
\end{figure}

\subsection{Perturbation \drt{R}esults}\label{sec:num_pert}

We first provide numerical validation for the first-order approximation given by Eq.~\eqref{eq:first_VNE}, which estimates how a network modification encoded by the perturbed Laplacian matrix $\Delta L$  affects  VNE. We created a random ER network with $N=1000$ nodes and $M=50,000$ edges, and subjected it to iterative uniform rewiring. We denote the original network $\drt{G^{(0)}}$ and the network after $t$ steps of uniform rewiring by $\drt{G^{(t)}}$, and we use $L^{(t)}$ and $h_t= h(G^{(t)})$ to denote the respective Laplacian matrices and VNEs for each $t=0,1,2,\dots$.  In Fig.~\ref{fig:VNE_perturb_1}(a), we compare the true values of $\{h_t \}$ of the rewired network with predicted values using the first-order approximation given by Eq.~\eqref{eq:first_VNE} for $K=1$ trial of uniform rewiring. These are in very good agreement for small $t$. We point out that the first-order approximation is expected to improve in accuracy as the eigenvalues $\{\lambda_i\}$ become larger, which typically occurs as $N$ and $M$ increase. We note that the first-order approximations described in Sec.~\ref{sec:Cont} can become inaccurate when $N$ and $M$ are too small.

In the next experiment, we support the results of Sec.~\ref{sec:Expected} in which we analyze the expected changes $\mathbf{E}[\Delta L]$ and $\mathbf{E}[\Delta h]$ under uniform rewiring. We created an ER network with $N=100$ nodes and $M=1,000$ edges, and subjected it to $K=10,000$ trials of iterative  uniform rewiring.  In Fig.~\ref{fig:VNE_perturb_1}(b), we compare the empirical mean ${\mathbf{E}}[\Delta L_{ij}^{(t)}]$ (symbols) to its expectation  given by Theorem \ref{thm:expected} (dashed lines). We make the comparison for three entries: $\Delta L_{ii}$ for a diagonal entry, $\Delta L_{ij}$ for an entry in which $(i,j)$ is an edge in $\drt{G^{(0)}}$, and $\Delta L_{ij}$ for an entry in which $(i,j)$ is not an edge in $\drt{G^{(0)}}$. Error bars indicate standard error.

\begin{figure}[t]
\centering
\includegraphics[width=\linewidth]{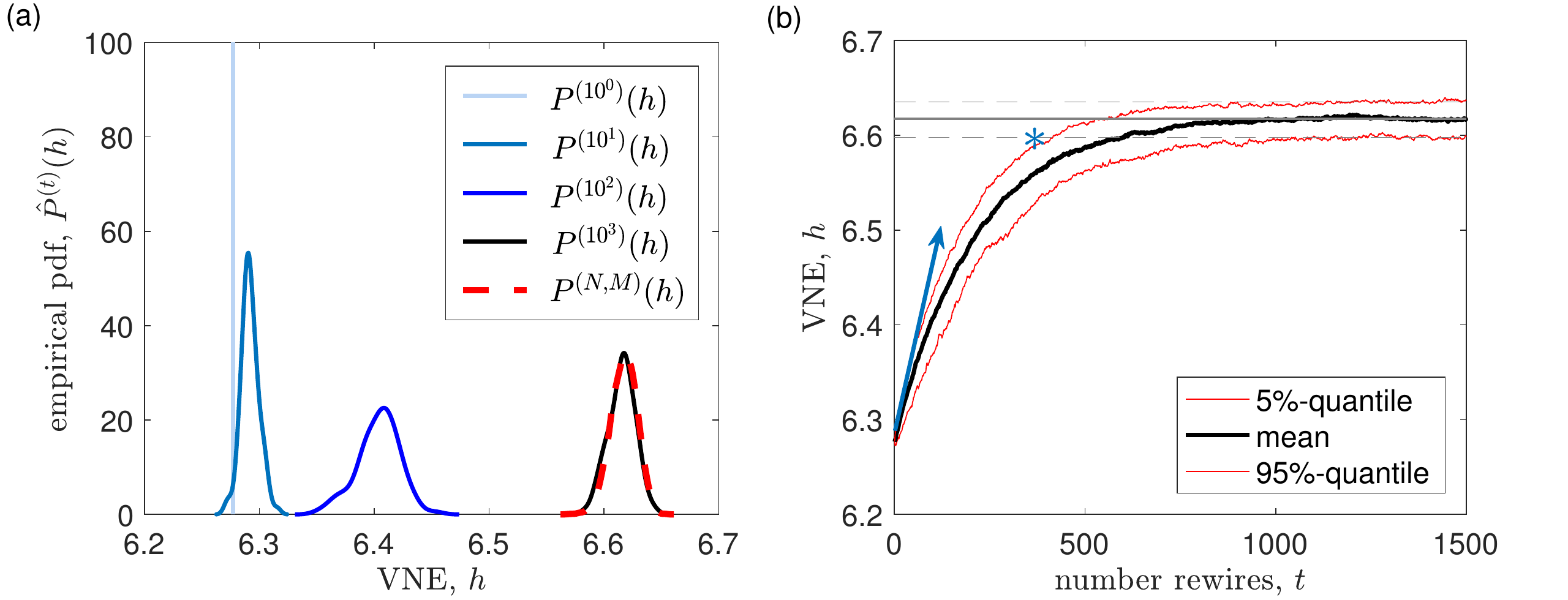} 
\caption{
Evolution of empirical distribution $ {P}^{(t)}(h)$ of VNE for a word-adjacency network \cite{adjnoun} subjected to \drt{stochastic} uniform rewiring. 
(a) We compare $ P^{(t)}(h)$ for $t\in\{0,10,100,1000\}$ to an empirical VNE distribution $P^{(N,M)}(h)$ for the ER random-network ensemble with identical $N$ and $M$. As $t\to\infty$,  $ P^{(t)}(h)$ evolves from a Dirac delta function  $ \delta_{h_0}(h)$ at $t=0$ to $ {P}^{(N,M)}(h)$. 
(b) We compare the 5\% quantile, \drt{mean}, and 95\% quantile for $ P^{(t)}(h)$ to those of $ P^{(N,M)}(h)$ (horizontal lines).
The blue arrow  indicates the slope $\mathbf{E} [h'(0)]$ which we approximate by Eq.~\eqref{eq:expected_eigen_perturb}. The blue star indicates the intersection between \drt{the $\alpha=5\%$ quantile of $ P^{(N,M)}(h)$ and an extrapolation that has initial slope  $\mathbf{E} [h'(0)]$ and converges to the mean $\overline{h}^{(N,M)}$ of $ {P}^{(N,M)}(h)$ such that the difference exponentially decays.}  In Sec.~\ref{sec:num_empirical}, we define an efficient quantity for network-ensemble comparison, $B_\alpha$, based on this \drt{exponential extrapolation}.
}
\label{fig:entropy_rewire}
\end{figure}

\subsection{Distributional Convergence of VNE}\label{sec:num_conv}

In Sec.~\ref{sec:CMT}, we showed that the distribution of VNEs across the ensemble of networks obtained after iterative stochastic uniform and degree-preserved rewires converges, respectively, to the distribution of VNEs across the ER and configuration random-network ensembles. 
Here, we support this result by studying uniform  rewiring for an empirical network: an adjacency network of words in the novel David Copperfield by Charles Dickens \cite{adjnoun}. The network contains $N = 112$ nodes (which represent the adjectives and nouns with highest frequency in the book) and $M=425$ edges (which represent a pair of words that occur adjacent to one another). 

We study how stochastic rewiring affects the VNE of this network by considering the distribution $P^{(t)}(h)$ of VNEs across networks $\drt{G^{(t)}}$ obtained after $t$ rewires.  In Fig.~\ref{fig:entropy_rewire}(a), we show by solid curves the empirical distributions $ P^{(t)}(h)$ for several values of $t\in\{0,10,100,1000\}$. The distributions are estimated using $K=1,000$ trials of rewiring for each $t$. Note that at time $t=0$, $ P^{(0)}(h) = \delta_{h_0}(h)$ is a Dirac delta function at $h_0=h(G^{(0)})\doteq 6.277$. As $t$ increases, $ P^{(t)}(h)$ widens and shifts to the right and eventually converges to $P^{(N,M)}(h)$, the distribution of VNE for the corresponding ER ensemble $\mathbf{G}_{N,M}$ (estimated using $K=10,000$ sample ER networks and shown by the dashed curve).

In Fig.~\ref{fig:entropy_rewire}(b), we further study the convergence of $ P^{(t)}(h) \to P^{(N,M)}(h) $ by plotting the 5\% quantile, \drt{mean} and 95\% quantile of $ {P}^{(t)}(h)$. These respectively converge to the  5\% quantile, mean and 95\% quantile for the distribution $P^{(N,M)}(h)$. The horizontal solid line indicates the mean VNE across $\mathbf{G}_{N,M}$ given by
\begin{equation}
\overline{h}^{(N,M)} = \int_0^\infty hP^{(N,M)}(h') dh'  . \label{eq:MEAN}
\end{equation}
We define the $\alpha$-quantile of $P^{(N,M)}(h) $ by
\begin{equation}
\drt{H^{(N,M)}}{(\alpha)} = H~\text{ such that }~   \alpha = \int_{0}^H P^{(N,M)}(h')dh',\label{eq:quant}
\end{equation}
and we plot $\drt{H^{(N,M)}}(0.05)$ and $\drt{H^{(N,M)}}(0.95)$ by horizontal dashed lines. 
\drt{
These quantiles were numerically approximated for $K=10,000$ sample ER networks. Numerically estimating $H^{(N,M)}(\alpha)$ is obviously associated with a computational cost, which depends on the accuracy required by the application. We note, however, that any network-ensemble comparison using network summary statistics  requires knowledge about how the summary statistic varies across the network ensemble. This highlights the need for further theory development for the distribution of VNE (and other summary statistics) across network ensembles.
}

%

\drt{Returning to Fig.~\ref{fig:entropy_rewire}(b),} because $h(G^{(0)}) \doteq 6.277$ is much smaller than the typical VNE values for the ensemble, the empirical network is much more irregular than is typical for the ensemble. Moreover, one can observe in Fig.~\ref{fig:entropy_rewire}(b)  that the distribution $P^{(t)}(h)$ obtained after $t=1000$ uniform rewires closely resembles the distribution $P^{(N,M)}(h)$. We can therefore conclude from Fig.~\ref{fig:entropy_rewire} that the word-adjacency network is atypical for the ER random-network ensemble.  In the next section, we more rigorously describe how to use the VNE of stochastically rewired networks to study and quantify  network-ensemble comparisons.

\subsection{Network-Ensemble Comparisons for Empirical Networks}\label{sec:num_empirical}


Given the convergence $P^{(t)}(h)\to P^{(N,M)}(h)$, there are many different ways to define and quantify what it means for a network to ``resemble a typical network.''  For example,  one could ask how many rewires are necessary for $\overline{h}^{(t)}$, the mean VNE of a network obtained after $t$ rewires, to be within some range of the ensemble mean, $\overline{h}^{(N,M)}$. Or one could measure the smallest $t$ such that $\overline{h}^{(t)}\in[\drt{H^{(N,M)}}(\alpha),\drt{H^{(N,M)}}(1-\alpha)]\subset \mathbb{R}$, where $\drt{H^{(N,M)}}(\alpha)$ is the $\alpha$-quantile given by Eq.~\eqref{eq:quant}. Another possibility is to ask how many rewires are necessary (on average) for $h_t=h(G^{(t)})$ of a rewired network $G^{(t)}$ to first fall within this range---that is, the mean hitting time 
\begin{equation}
\tau_\alpha = \mathbf{E}\left[ \min_t~\{t: h_t\in[\drt{H^{(N,M)}}(\alpha),\drt{H^{(N,M)}}(1-\alpha)]\} \right]. \label{eq:tau}
\end{equation}
Given that a stochastic rewiring process can be modeled as a random walk on a set of networks, $\tau_\alpha$ is equivalent to the mean first-passage time for a random walk that starts at network $\drt{G^{(0)}}$ and reaches the subset of networks $\{ {G}_s\in\mathcal{G}_{N,M} :h(G_s)\in[\drt{H^{(N,M)}}(\alpha),\drt{H^{(N,M)}}(1-\alpha)]\}$.  Unfortunately, these methods are computationally expensive in that they require  one to simulate $t\gg1$ stochastic rewires across $K\gg1$ independent trials of rewiring, all while computing the VNE for the many rewired network realizations.

Thus motivated, we propose a computationally efficient technique for network-ensemble comparison that does not require computing the VNE of rewired networks. In fact, it avoids simulating stochastic rewires altogether. Instead, we introduce a quantity that utilizes our first-order perturbation analysis of Sec.~\ref{sec:Expected}.

\begin{definition}[\drt{Exponential Extrapolation for $\alpha$-Quantile Intersect}]\label{def:BB}
\drt{Suppose we fit an exponential model to $h_t$, which has the following form
\begin{equation}\label{eq:expo_2}
\tilde{h}(t) = \overline{h}^{(N,M)} - (\overline{h}^{(N,M)} - h_0) \exp \left( -\frac{\mathbf{E} [h'(0)]}{\overline{h}^{(N,M)} - h_0} t \right)
\end{equation}
The model satisfies the conditions that when $t = 0$ we have $\tilde{h}(0) = h_0$, $\tilde{h}'(0) = \mathbf{E} [h'(0)]$, and  $\tilde{h}(t) \to \overline{h}^{(N,M)}$ as $t\to\infty$.

By solving the equation
\begin{equation}
\tilde{h}({B_\alpha}) = \drt{H^{(N,M)}}(\alpha)
\end{equation}
we obtain
\begin{equation}\label{eq:expo_3}
B_\alpha = - \log \left( \frac{\overline{h}^{(N,M)} - \drt{H^{(N,M)}}(\alpha)}{\overline{h}^{(N,M)} - h_0} \right) \frac{\overline{h}^{(N,M)} - h_0}{\mathbf{E} [h'(0)]}
\end{equation}
where $\drt{H^{(N,M)}}{(\alpha)}$ is given by Eq.~\eqref{eq:quant}, $\overline{h}^{(N,M)}$ is given by Eq.~\eqref{eq:MEAN} and $\mathbf{E} [h'(0)]$ is given by Eq.~\eqref{eq:expected_eigen_perturb}. Note that for $B_\alpha$ to be properly defined, $(\overline{h}^{(N,M)} - \drt{H^{(N,M)}}(\alpha))$ must have the same sign as $(\overline{h}^{(N,M)} - h_0)$. For example, if $h_0 < \overline{h}^{(N,M)}$, then $\alpha$ could be $0.05$ but not $0.95$.}
\end{definition}

The quantity $B_\alpha$ is \drt{an exponential} extrapolation that estimates the number $t$ of uniform rewires required to modify a given network so that $\drt{h_t}$ falls between the $\alpha$ and $(1-\alpha)$ quantiles of $P^{(N,M)}(h)$. For example, returning to the experiment described in Sec.~\ref{sec:num_conv}, the blue arrow in Fig.~\ref{fig:entropy_rewire}(b) indicates the \drt{initial slope $\mathbf{E} [h'(0)]$},
and the blue star indicates the intersection point $(B_\alpha,\drt{H^{(N,M)}}(\alpha))$ for $\alpha=0.05$ \drt{at which $\tilde{h}(B_\alpha) = \drt{H^{(N,M)}}(\alpha)$}.

\begin{table}[h]
\centering
    \begin{tabular}{c c c c}
    \hline\hline
    Network & $N$ & $M$ & Reference \\ \hline
    dolphin social network & 62 & 159 & \cite{dolphins} \\ 
    Les Mis\'erables characters & 77 & 254 & \cite{lesmis} \\ 
    word adjacency in David Copperfield & 112 & 425 & \cite{adjnoun} \\ 
    jazz collaborations & 198 & 2742 & \cite{jazz} \\
    C. elegans neuronal network & 297 & 2148 & \cite{neural} \\ 
    C. elegans metabolic network & 453 & 2025 & \cite{metabolic} \\ 
   world airport network & 500 & 2980 & \cite{airport} \\ 
       Caltech Facebook network & 762  & 16651 & \cite{traud2011comparing} \\
       university email messages  & 1133 & 5451 & \cite{email} \\
    U. S. power grid & 4941   & 6594 & \cite{neural} \\
    \hline\hline
    \end{tabular}
\caption{Empirical networks studied in Fig.~\ref{fig:entropy_comparison}.}
\label{table1}
\end{table}

\begin{figure}[t] 
\centering
\includegraphics[width=\linewidth]{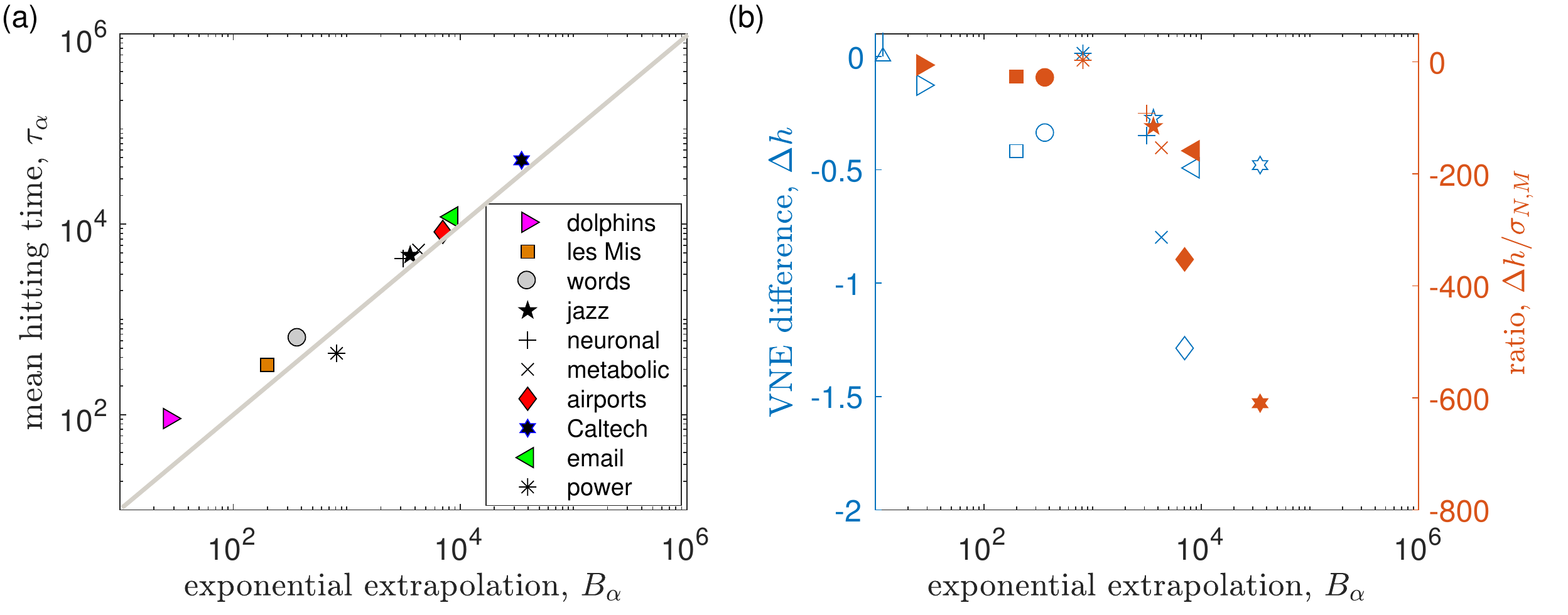}
\caption{ 
Network-ensemble comparisons for the empirical networks in Table~\ref{table1} \drt{and the associated ER ensembles}.
(a)~Observed mean hitting times $\tau_\alpha$ given by Eq.~\eqref{eq:tau} strongly correlate with the \drt{exponential} extrapolation $B_\alpha$ given by Eq.~\eqref{eq:expo_3} (shown with $\alpha=0.05$).
(b)~We compare  $B_\alpha$ to the difference $ \Delta h= h(G^{(0)}) - \overline{h}^{(N,M)}$ between the VNE of the empirical networks, $ h(G^{(0)})$, and the mean VNE across the appropriate ER ensembles, $\overline{h}^{(N,M)}$ (blue symbols, left vertical axis) as well as to the ratio of $\Delta h$ to the standard deviation $\sigma_{N,M}$ of $P^{(N,M)}(h)$ (red symbols, right vertical axis).
} 
\label{fig:entropy_comparison}
\end{figure}

\begin{figure}[t] 
\centering
\includegraphics[width=\linewidth]{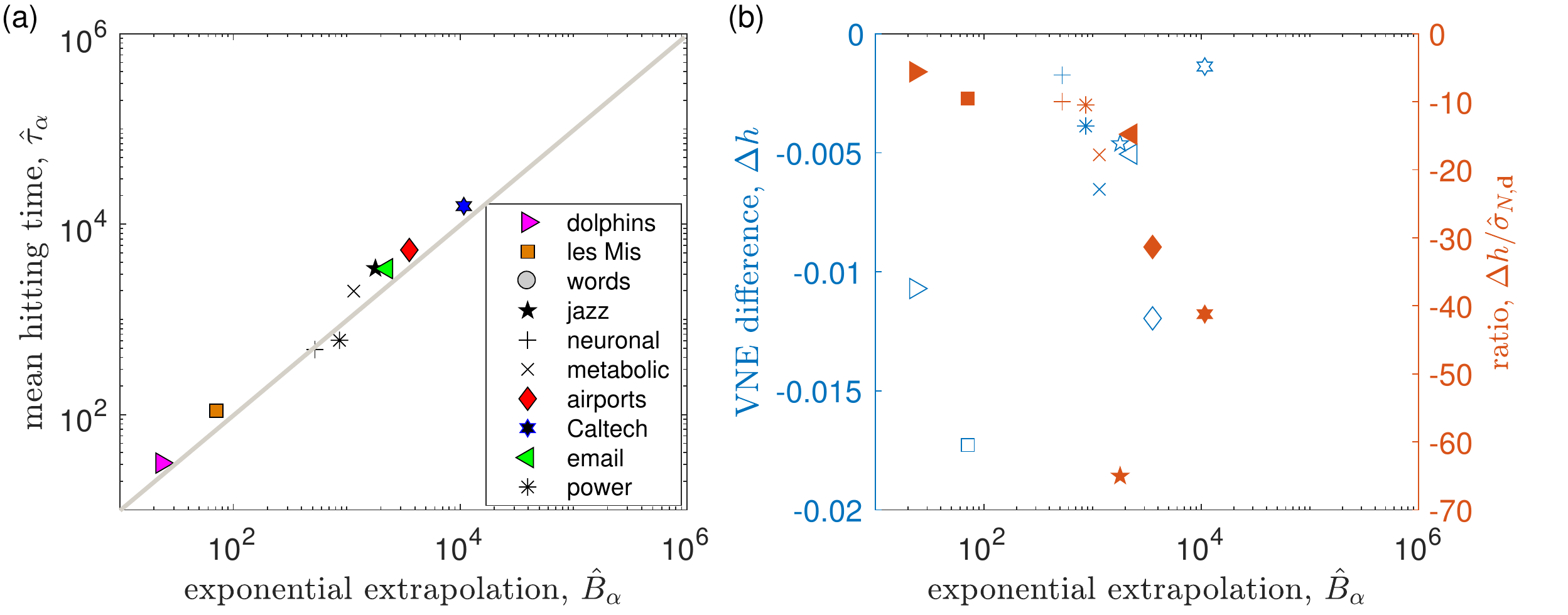}
\caption{
\drt{Network-ensemble comparisons for the empirical networks in Table~\ref{table1} and the configuration model.
(a)~Observed mean hitting times $\hat{\tau}_\alpha$  strongly correlate with the exponential extrapolation $\hat{B}_\alpha$ given by Eq.~\eqref{eq:expo_3} (shown with $\alpha=0.05$), which uses a numerical estimate for ${\bf E}[h'(0)]$.
(b)~Comparison of $\hat{B}_\alpha$ to  $ \Delta h= h(G^{(0)}) - \overline{h}^{(N,{\bf d})}$ 
(blue symbols, left vertical axis) and to the ratio of $\Delta h/ \hat{\sigma}_{N,{\bf d}}$ (red symbols, right vertical axis), where $\hat{\sigma}_{N,{\bf d}}$ is the standard deviation of $\hat{P}^{(N,{\bf d})}(h)$.
} }
\label{fig:entropy_comparison2}
\end{figure}

We now study network-ensemble comparisons for the empirical networks described in Table~\ref{table1}. In Fig.~\ref{fig:entropy_comparison}(a), we compare $B_\alpha$ given by Eq.~\eqref{eq:expo_3} to the mean first-passage time $\tau_\alpha$ given by Eq.~\eqref{eq:tau} \drt{for the associated ER ensemble}. For each empirical network, we perform $K = 10$ independent simulations of repeated uniform rewiring, which we iterate until $h_t\in[H(\alpha),H(1-\alpha)]$ (the ``hitting'' criterion). 
The $\alpha$-quantiles of $P^{(N,M)}(h)$ are estimated using $K=100$ samples from the ER random-network ensemble. We observe a strong linear correlation between $B_\alpha$ and $\tau_\alpha$.  In fact, the results fall along a diagonal line \drt{indicating $\tau_\alpha=B_\alpha$.}
%
In Fig.~\ref{fig:entropy_comparison}(b), we compare $B_\alpha$ to two other quantities: (i) the difference $\Delta h = h(G^{(0)}) - \overline{h}^{(N,M)}$ in VNE between the original network and the ensemble mean $\overline{h}^{(N,M)}$ as well as (ii)~the ratio of $\Delta h$ to the standard deviation $\sigma_{N,M}$ of $P^{(N,M)}(h)$. Interestingly, these two other quantities do not as strongly correlate with $B_\alpha$ and $\tau_\alpha$.
\drt{Finally, we note that the result for the power grid in Fig.~\ref{fig:entropy_comparison} does not use the first-order approximation for ${\bf E}[h'(0)]$ given by Eq.~\eqref{eq:expected_eigen_perturb}. We find that it is not accurate for this network, and in fact has the wrong sign, which causes the exponential extrapolation given by Eq.~\eqref{eq:expo_2} to diverge rather than converge. (Interestingly, the power grid is the only network in Table \ref{table1} for which $h_0>\overline{h}^{(N,M)}$ and ${\bf E}[h'(0)]<0$.) Therefore, for this network we used a numerical estimate for ${\bf E}[h'(0)]$ based on the  of VNE change $(h_1 - h_0)$ due to a single uniform rewire, which we averaged across 1000  simulations (see discussion in Sec.~\ref{sec:Roadmap}).}

\drt{

In Fig.~\ref{fig:entropy_comparison2}, we present similar results as those shown in Fig.~\ref{fig:entropy_comparison}, except we now compare the empirical networks from Table~\ref{table1} to configuration-model ensembles containing networks with the same degree sequence. We estimated $\hat{H}^{(N,{\bf d})}(\alpha)$---which is given by Eq.~\eqref{eq:quant} under the variable substitution $\hat{P}^{(N,{\bf d})}(h) \mapsto P^{(N,M)}(h)$---by sampling the  configuration-model ensemble using degree-preserved rewiring. Starting with the associated empirical network, we implemented $10^5$ degree-preserved rewires. Then, we implemented $10^5$ additional rewires, sampling the network's VNE every $100$ rewires, which allowed us to estimate $\hat{P}^{(N,{\bf d})}(h)$ and $\hat{H}^{(N,{\bf d})}(\alpha)$ using $K=1000$ VNE samples.
To numerically estimate ${\bf E}[h'(0)]$ (see  Sec.~\ref{sec:Roadmap}), we simulated $1000$ degree-preserved rewires, each time modifying the original network by a single rewire, and computed mean difference $(h_1 - h_0)$ across the simulations. Using this estimate for ${\bf E}[h'(0)]$, we obtained an exponential extrapolation for the intersect of $h_t$ with $\hat{H}^{(N,{\bf d})}(\alpha)$ using Eq.~\eqref{eq:expo_3} with the variable substitutions  $\overline{h}^{(N,{\bf d})} \mapsto \overline{h}^{(N,{M})}$ and $\hat{H}^{(N,{\bf d})} \mapsto H^{(N,{M})}$.

In Fig.~\ref{fig:entropy_comparison2}(A), we compare    $\hat{B}_\alpha$ to an empirical mean hitting time $\hat{\tau}_\alpha$ given by Eq.~\eqref{eq:tau} under the variable substitution $\hat{H}^{(N,{\bf d})} \mapsto H^{(N,{M})}$.
%
In Fig.~\ref{fig:entropy_comparison2}(B), we show that  $ \Delta h= h(\hat{G}^{(0)}) - \overline{h}^{(N,{\bf d})}$  and $\Delta h / \hat{\sigma}_{N,{\bf d}}$, where $\hat{\sigma}_{N,{\bf d}}$ is the empirical standard deviation of $\hat{P}^{(N,{\bf d})}(h)$,  do not strongly correlate with $\hat{B}_\alpha$ (or $\hat{\tau}_\alpha$).
%
It can observed that the empirical networks generally require fewer stochastic rewires to obtain typical VNE values for the configuration-model ensembles than was required for the ER ensembles (recall Fig.~\ref{fig:entropy_comparison}).  Interestingly, the word-adjacency network  \cite{adjnoun} (see also Fig.~\ref{fig:entropy_rewire}) is omitted from Fig.~\ref{fig:entropy_comparison2}(A), because it was found to lie within the $\alpha=0.05$ and $\alpha=0.95$ quantiles of $\hat{P}^{(N,{\bf d})}(h)$, indicating that it closely resembles a typical configuration-model graph according to the VNE statistic. 

}

\section{Discussion}\label{sec:Discuss}

We have studied the von Neumann Entropy (VNE) of networks subjected to two stochastic rewiring processes: uniform and degree-preserved rewiring. We presented our main mathematical results in Sec.~\ref{sec:Main}. First, we proved that the network-ensemble given by networks obtained through iterative uniform rewiring converges to the Erd\H{o}s-R\'enyi ensemble $\mathbf{G}_{N,M}$ of simple networks.  Next, we proved that the distribution of network summary statistics for networks obtained from iterative uniform and degree-preserved rewiring  converge to their respective distributions associated with the Erd\H{o}s-R\'enyi $\mathbf{G}_{N,M}$ and configuration $\drt{\hat{\mathbf{G}}}_{N,\bf{d}}$ ensembles (offering insight toward network-ensemble comparisons). We also conducted a perturbation analysis for how rewiring affects VNE, offering insight toward network-network comparisons. In particular, we obtained a first-order approximation for the expected change in VNE after $t$ uniform  rewires.

In Sec.~\ref{sec:Num}, we showed that the study of VNE for an empirical network subjected to repeated uniform rewires allows one to assess how many rewires are required before the rewired networks obtain VNE values that are typical for the $\mathbf{G}_{N,M}$ \drt{and  $\hat{\mathbf{G}}_{N,{\bf d}}$ ensembles}. Importantly, such a numerical study can be computationally infeasible since it can require simulating many steps of rewiring, many independent trials of rewiring, and repeated calculations of VNE for the rewired networks. Thus motivated,  we introduced a computationally efficient quantity $B_\alpha$ to quantify network-ensemble comparisons. It combines our perturbation and convergence analyses and is based on \drt{an exponential} extrapolation for when the VNE of rewired networks intersects an $\alpha$ quantile of the VNE distribution $P(h)$ for the appropriate ensemble (see Fig.~\ref{fig:entropy_rewire}). The quantity $B_\alpha$ is computationally efficient since it does not require \drt{iteratively} rewiring a network nor recomputing VNE for these networks.  
\drt{In the case of $\mathbf{G}_{N,M}$, $B_\alpha$ used an analytical approximation for the perturbative influence of uniform rewiring on VNE (see Corr.~\ref{thm:first}). In the case of $\hat{\mathbf{G}}_{N,{\bf d}}$,  $\hat{B}_\alpha$ required a  numerical estimate for the perturbative influence of degree-preserved rewiring on VNE, 
which provides a road map for how to extend this} methodology for 
other random-network ensembles \cite{boccaletti2014structure,holme2012temporal,newman2010networks} and  \drt{other} stochastic rewiring processes  \cite{fosdick2016configuring}.
\drt{Because the approach of linearizing the effect of stochastic rewires to estimate network-ensemble comparisons is computationally efficient, future work should explore its application to model selection and hypothesis testing \cite{bassett2013robust,caceres2016model,kolaczyk2014statistical,milo2002network}.}

To our knowledge, this is the first use of VNE for network-ensemble comparison. 
We have focused  on VNE due to growing interest in VNE-based network-network comparisons, such as clustering network layers in multilayer networks \cite{de2015structural,de2016spectral}. We point out, however, that our mathematical techniques\drt{---specifically, the approach of linearizing the effect of rewiring so as to obtain an exponential extrapolation $B_\alpha$---}can be extended to study convergence and assess network-ensemble comparisons through other network summary statistics (e.g., degree distribution, size, clustering coefficient, and so on). For example, it would be interesting to extend our work to a complementary definition for VNE that was  recently introduced  \cite{de2016spectral}.
\drt{Because one can numerically approximate the linear effect of stochastic rewiring on a summary statistic (e.g., see the results in Fig.~\ref{fig:entropy_comparison2} and discussion in Sec.~\ref{sec:Roadmap}), our approach should be widely generalizable to other network summary statistics as well as vectors of  statistics. This approach may  shed light toward the relation between different summary statistics  and identify which ones are more meaningful to different network ensembles, whether it's basic statistics, information-theoretic measures such as VNE (e.g., Eq.~\eqref{eq:vne1} or otherwise  \cite{de2016spectral}), or some new summary statistic yet to be discovered.}

\appendix

\section{Proof of Theorem \ref{thm:CONVERGENCE}} \label{appendixA0}
\begin{proof}
The result follows from showing the Markov chain is connected, aperiodic and degree regular. 

We first prove the Markov chain described in Eq.~\eqref{eq:NDP_chain} corresponds to a connected graph. To this end, we will show for any two networks there exists a path---that is, a sequence of edge swaps allowed by uniform rewiring---between the two networks.
Let $G_s = ( \mathcal{V},\mathcal{E}_{s})\in\mathbf{G}_{N,M}$ and $G_r = ( \mathcal{V},\mathcal{E}_{r})\in\mathbf{G}_{N,M}$ and define $\Delta^{(s)} = \mathcal{E}_{s} \setminus \mathcal{E}_{r}$ and $\Delta^{(r)} = \mathcal{E}_{r} \setminus \mathcal{E}_{s}$ indicate, respectively, the set of edges in $\mathcal{E}_{s}$ and $\mathcal{E}_{r}$ that are not in the other edge set. Because $M=|\mathcal{E}_{s}| = |\mathcal{E}_{r}| $, it follows that $T = |\Delta ^{(s)}| = |\Delta^{(r)}|$. We enumerate the entries in $\Delta^{(s)}$ and $\Delta^{(r)}$ as $\Delta^{(s)}_j$ and  $\Delta^{(r)}_j$ for $j=\{1,\dots,T\}$ and define the family of maps $T_{r,s}^{(j)}: \mathbf{G}_{N,M}\mapsto \mathbf{G}_{N,M}$ by $( \mathcal{V},\mathcal{E}) \mapsto \big( \mathcal{V},( \mathcal{E}\setminus \Delta_j^{(r)})\cup \Delta_j^{(s)}\big). $ It follows  that 
\begin{equation}
T_{r,s}^{(1)} (T_{r,s}^{(2)}(\dots T_{r,s}^{(T)}(G_r))) = G_s.
\end{equation}
 We can similarly define $T_{s,t}^{(j)}$ so that 
 \begin{equation}
 T_{s,r}^{(1)} (T_{s,r}^{(2)}(\dots T_{s,r}^{(T)}(G_s))) = G_r.
 \end{equation}

Next, we prove the graph is degree-regular. Consider a network $G_s\in\mathcal{G}_{N,M}$ containing $N$ nodes and $M$ edges. It follows that there are $M$ possibilities for edge removal and $N(N-1)/2 - M +1$ possibilities for new edges to add. (Here, the $+1$ allows for the removed edge to be replaced.) Moreover, for any transition $G_s\to G_r$, there exists a transition $G_r\to G_s$ with identical rate. Therefore, the Markov chain corresponds to an undirected network in which all nodes have degree $d_{U} = M(N(N-1)/2 - M +1)$. 

Finally, we prove aperiodicity. Because the graph is connected and contains self-edges, the Markov chain is aperiodic.

\end{proof}

\section{Proof of Theorem \ref{thm:main1}} \label{appendixA}
\begin{proof}
Taylor expansion near $\epsilon=0$ gives 
\begin{equation}  
h( \epsilon) = h(0) + \epsilon h'(0) + \mathcal{O}(\epsilon^2).  \nonumber
\end{equation}
Here we show that $h'(0)$ is given by Eq.~\eqref{eq:vne_perturb}. Using Eq.~\eqref{eq:vne2}, the VNE entropy of a network corresponding to Laplacian matrix $L(\epsilon)$ is given by
	\begin{align}
		h(\epsilon)  &= - \sum_{i } f \left( {\lambda_i(\epsilon)} \right) ,
	\end{align}
where
	\begin{align}
		f(x)=\frac{x}{2M}\log_2\left(\frac{x}{2M}\right)
	\end{align} 
has derivative
	\begin{align}
		\frac{df}{dx} &= \frac{1}{2M} \left(\log_2\left(\frac{x}{2M}\right) +   \frac{1}{\log(2)}\right). \nonumber
	\end{align}
Using the linearity property of differentiation, we express the derivative of $h(\epsilon)$ via partial derivatives as
	\begin{align}
		\frac{dh}{d\epsilon} &= \sum_i \frac{\drt{df}}{d\lambda}\frac{d\lambda}{d\epsilon}.
	\end{align}
Letting $\epsilon=0$, we find
	\begin{align}
		h'(0)  = \left. \frac{dh(\epsilon)}{d\epsilon} \right|_{\epsilon=0} 
			&= \drt{-} \sum_i  \lambda'(0) \left[\frac{1}{2M} \left(\log_2\left(\frac{\lambda_i}{2M}\right) +   \frac{1}{\log(2)}\right) \right] .
\end{align}
We substitute $ \lambda'(0)=\bm v^{(i)})^T \Delta  {L} \bm v^{(i)}$ from Eq.~\eqref{eq:eigen_perturb_b} to obtain Eq.~\eqref{eq:vne_perturb}.
\end{proof}

\section{Proof of Corollary \ref{cor:1edge}} \label{appendixB}
\begin{proof}
For unweighted networks, all non-diagonal entries $L_{ij}$ in  are either 0 (if there is no edge) or -1 (if there is an edge $(i,j)\in\mathcal{E}$). The addition of an edge $(p,q)$ implies $L_{pq}=L_{qp}=-1$, and because $\sum_i L_{ij}=0$ by definition, $L_{ii} = \sum_{j\not=i} L_{ij}$ and any perturbation of off-diagonal elements must be reflected in the diagonal elements. Consideration of an edge removal leads to an analogous result, albeit with an opposite sign, and therefore $\Delta L$ must be of the form given by Eq.~\eqref{eq:DL}.  It is straightforward to show 
\begin{align}
	(\bm v^{(n)})^T \Delta  {L}^{(pq)} \bm v^{(n)} &= (\bm v^{(n)}_p-\bm v^{(m)}_q)^2.
\end{align}
We substitute this result into Eq.~\eqref{eq:vne_perturb} to obtain Eq.~\eqref{eq:vne_perturb2}.
\end{proof}

\section{Proof of Theorem \ref{thm:expected}} \label{appendixC}
\begin{proof}
The process of randomly rewiring an edge $(p,q)$ to $(r,s)$ can be decomposed into two steps. The first step is removing an edge $(p,q)$ from the original graph $G^{(0)}$, resulting in an intermediate graph $G^{(1)}$. The second step is adding an edge $(r,s)$ to the graph $G^{(1)}$, resulting in the rewired graph $G^{(2)}$. Let $L^{(0)}$ denote the Laplacian matrix of the original graph $G^{(0)}$, $L^{(1)}$ denote the Laplacian matrix of the intermediate graph $G^{(1)}$, and $L^{(2)}$ denote the Laplacian matrix of the rewired graph $G^{(2)}$, then we have $L^{(1)} = L^{(0)} + \Delta L^{(0)}$, $L^{(2)} = L^{(1)} + \Delta L^{(1)}$. In terms of our previous notations, we have
\begin{equation}
L = L^{(0)} \text{ , }  L' = L^{(2)}  \text{ , } \Delta L = \Delta L^{(0)} + \Delta L^{(1)}
\end{equation}

\subsection{Removing an edge}

Since removing an edge $(p,q)$ means $A_{pq}$ and $ A_{qp}$ change from $1$ to $0$, the elements of $\Delta L^{(0)}_{ij}$ are given by
\begin{equation}
\Delta L^{(0)}_{ij} = \begin{cases} 
-1 &\mbox{if } i = j \in \{ p,q \} \\
1 &\mbox{if } i \in \{ p,q \} \text{ and } j \in \{ p,q \} \setminus i \\
0 &\mbox{otherwise} .
\end{cases} 
\end{equation}
Using that edges are removed uniformly at random, the expected values of $\{\Delta L^{(0)}_{ij}\}$ are given by
\begin{equation}\label{eq:Exp1}
\mathbf{E} [ \Delta L^{(0)}_{ij}]= \begin{cases} 
P(p = i \text{ or } q = i) \times (-1) &\mbox{if } i = j  \\
P((p = i \text{ and } q = j) \text{ or } (p = j \text{ and } q = i)) \times 1 &\mbox{if } i \neq j 
\end{cases} 
\end{equation}
Since there are $M$ edges in total, and we can only remove an edge when $A_{ij}=A_{ji}=1$, we can write down the probabilities as
\begin{equation}
P(p = i \text{ or } q = i) = \frac{d_i}{M}
\end{equation}
and 
\begin{equation}
P((p = i \text{ and } q = j) \text{ or } (p = j \text{ and } q = i)) = \frac{A_{ij}}{M}.
\end{equation}
We substitute these probabilities into Eq.~\eqref{eq:Exp1} to obtain
\begin{equation} \label{eq:remove}
\mathbf{E} [ \Delta L^{(0)}_{ij}]= \begin{cases} 
- \frac{d_i}{M} &\mbox{if } i = j \\
\frac{A_{ij}}{M} &\mbox{if } i \neq j 
\end{cases} 
\end{equation}

\subsection{Adding an edge}

Since adding an edge $(r,s)$ means $A_{rs} $ and $ A_{sr}$ change from $0$ to $1$, the elements  $\{\Delta L^{(1)}_{ij}\}$ are given by
\begin{equation}
\Delta L^{(1)}_{ij} = \begin{cases}
1 &\mbox{if } i = j \in \{ r,s \} \\ 
-1 &\mbox{if } i \in \{ r,s \} \text{ and } j \in \{ r,s \} \setminus i \\
0 &\mbox{otherwise} ,
\end{cases} 
\end{equation}
which have the expectations
\begin{equation}\label{eq:Exp2}
\mathbf{E} [ \Delta L^{(1)}_{ij}]= \begin{cases} 
P(r = i \text{ or } s = i) \times 1 &\mbox{if } i = j  \\
P((r = i \text{ and } s = j) \text{ or } (r = j \text{ and } s = i)) \times (-1) &\mbox{if } i \neq j .
\end{cases} 
\end{equation}
Since there are $\frac{N(N-1)}{2}$ possible edges in total for a graph with $N$ nodes, and we can only add an edge when $A_{ij}=A_{ji}=0$, and $\{i,j\} \neq \{p,q\}$. Therefore, there are $R =\frac{N(N-1)}{2} - M +1$ possible edges to add,  yielding the probabilities 
\begin{equation}
P(r = i \text{ or } s = i) = \frac{N - 1- d_i}{R}
\end{equation}
and 
\begin{equation}
P((r = i \text{ and } s = j) \text{ or } (r = j \text{ and } s = i)) = \frac{1-A_{ij}}{R}.
\end{equation}
We substitute these probabilities into Eq.~\eqref{eq:Exp2} to obtain
\begin{equation} \label{eq:add}
\mathbf{E} [ \Delta L^{(1)}_{ij}]= \begin{cases}  \frac{N - 1- d_i}{R} &\mbox{if } i = j \\
-\frac{1-A_{ij}}{R} &\mbox{if } i \neq j .
\end{cases} 
\end{equation}

\subsection{Rewiring an edge}

By the linearity of expectation, we have
\begin{equation} \label{eq:linear}
\mathbf{E} [\Delta L]  = \mathbf{E} [ \Delta L^{(0)} ] + \mathbf{E} [ \Delta L^{(1)} ] .
\end{equation}
We substitute Eqs. \ref{eq:remove} and \ref{eq:add} into \ref{eq:linear} to obtain
\begin{equation} \label{eq:rewire}
\mathbf{E} [ \Delta L_{ij}]= 
\begin{cases}  \frac{N - 1- d_i}{\frac{N(N-1)}{2} - M+1} - \frac{d_i}{M} &\mbox{if } i = j\\
\frac{A_{ij}}{M} - \frac{1-A_{ij}}{\frac{N(N-1)}{2} - M+1} &\mbox{if } i \neq j .
\end{cases} 
\end{equation}
\end{proof}

\bibliographystyle{siamplain}
\bibliography{Entropy_bib}

\end{document}